\documentclass[a4paper,11pt,fleqn]{article}

\usepackage{jheppub}

\usepackage{cancel}

\usepackage{jheppub}

\usepackage{subfig}
\usepackage{amsmath}

\usepackage{amssymb}
\usepackage{latexsym}
\usepackage{wasysym}
\usepackage{pstricks}

\newcommand{\BIG}{All Order\ }

\newcommand{\hG}{\hat{G}}

\newcommand{\hF}{\hat{F}}

\newcommand{\hbe}{\hat{\beta}}

\newcommand{\hga}{\hat{\gamma}}

\newcommand{\ha}{\hat{a}}

\newcommand{\si}{{\, \mathrm{ si} }}

\newcommand{\bga}{\begin{gather}}

\newcommand{\bal}{\begin{align}}

\newcommand{\eal}{\end{align}}

\newcommand{\z}{\zeta}

\newcommand{\ice}[1]{{}}

\newcommand{\EQN}[1]{\label{#1}}

\newcommand{\ed}{\end{document}}

\newcommand{\prd}{\partial}
\newcommand{\ep}{\epsilon}
\newcommand{\beq}{\begin{equation}}
\newcommand{\eeq}{\end{equation}}
\newcommand{\bea}{\begin{eqnarray}}
\newcommand{\eea}{\end{eqnarray}}

\newcommand{\ba}{\begin{array}}
\newcommand{\ea}{\end{array}}

\newcommand{\als}{\boldmath{\alpha_s}}

\newcommand{\g}{\gamma}

\newcommand{\be}{\beta}
\newcommand{\bc}{\begin{center}}
\newcommand{\ec}{\end{center}}

\newcommand{\re}[1]{(\ref{#1})}
\newcommand{\unl}[1]{\underline{#1}}

\catcode`\@=11
\def\slash{\mathpalette\make@slash}
\def\make@slash#1#2{\setbox\z@\hbox{$#1#2$}%
  \hbox to 0pt{\hss$#1/$\hss\kern-\wd0}\box0}
\catcode`\@=12 

\def\bbuildrel#1_#2^#3%
{\mathrel{\mathop{\kern 0pt#1}\limits_{#2}^{#3}}}

\newcommand{\nnb}{\nonumber}

\newcommand{\MSbar}{\ensuremath{\overline{\text{MS}}}}

\newcommand{\Ghat}{\ensuremath{\widehat{\text{G}}}}

\def\beq{\begin{equation}}
\def\eeq{\end{equation}}
\def\bea{\begin{eqnarray}}
\def\eea{\end{eqnarray}}
\def\bq{\begin{quote}}
\def\eq{\end{quote}}

\def\nnb{\nonumber}

\def\nnb{\nonumber}

\def\ba{\begin{array}}
\def\ea{\end{array}}

\def\bbuildrel#1_#2^#3%
{\mathrel{\mathop{\kern 0pt#1}\limits_{#2}^{#3}}}

\newcommand{\epi}{\,\bbuildrel{=}_{}^{\pi}\,}

\newcommand{\epifour}{\,\bbuildrel{=\!\!\!=}_{}^{\pi^4}\,}

\newcommand{\episix}{\,\bbuildrel{=\!\!\!=}_{}^{\pi^6}\,}

\newcommand{\lQ}{\ell_\mu}

\newcommand{\defas}{\mathrel{\mathop:}=}
\newcommand{\mzv}[2][]{\zeta^{#1}_{#2 }}

\newcommand{\hZ}{\hat{Z}}
\newcommand{\hZa}{\hat{Z}_a}

\newcommand{\Za}{{Z}_a}

\title{
{
\boldmath
\!\!\!\!\!\!\!\!
The structure of  generic   anomalous dimensions
and no-$\pi$ theorem   for massless propagators 
}
}
\author[a]{P.~A.~Baikov,}
\author[b]{K. G. Chetyrkin}

\affiliation[a]{
Skobeltsyn Institute of Nuclear Physics, Lomonosov Moscow State University, 
1(2), Leninskie gory, Moscow  119991, Russian Federation
        }        
\affiliation[b]{
II Institut f\"ur Theoretische Physik,
Universit\"at  Hamburg, Luruper Chaussee 149, 22761 Hamburg, Germany
}

\emailAdd{baikov@theory.sinp.msu.ru}
\emailAdd{Konstantin.Chetyrkin@desy.de}

\abstract{

 Extending an argument of \cite{Baikov:2010hf} for the case of 5-loop massless
 propagators we prove a host of new exact model-independent relations between
 contributions proportional to odd and even zetas in generic \MSbar\ anomalous
 dimensions as well as in generic massless correlators. In particular, we find
 a new remarkable connection between coefficients in front of $\z_3$ and
 $\z_4$ in the 4-loop and 5-loop contributions to the QCD $\beta$-function
 respectively.  
It leads to a natural explanation of a
 simple mechanics behind mysterious cancellations of the $\pi$-dependent terms
 in one-scale Renormalization Group (RG) invariant Euclidean quantities
 recently discovered in \cite{Jamin:2017mul}.  We give a proof of this no-$\pi$
 theorem for a general case of (not necessarily scheme-independent) one-scale
 massless correlators.  All $\pi$-dependent terms in the {\bf six-loop}
 coefficient of an anomalous dimension (or a $\beta$-function) are shown to be
 explicitly expressible in terms of lower order coefficients for a general
 one-charge theory. For the case of a scalar $O(n)$ $\phi^4$ theory all our
 predictions for $\pi$-dependent terms in 6-loop anomalous dimensions are in
 full agreement with recent results of
 \cite{Batkovich:2016jus,Schnetz:2016fhy,Kompaniets:2017yct}.

}


\begin{document}


\maketitle

\section{Introduction \label{sec:intro}}

The seminal calculation of the Adler function at order $\alpha_s^3$
\cite{Gorishnii:1991vf} demonstrated for the first time a mysterious complete
cancellation of all contributions proportional to $\zeta_4$ (abounding in
separate diagrams) while odd zetas terms (that is those proportional to
$\zeta_3$ and $\zeta_5$) survived and appeared in the final result. Literally
the authors of \cite{Gorishnii:1991vf} wrote: {\em ``We would like to stress
  the cancellations of $\z_4$ in the final results for $R(s)$.  It is very
  interesting to find the origin of the cancellation of $\z_4$ in the physical
  quantity.''}

Since then it has been noted many times that {\em all} one scale physical
quantities are indeed free from even zetas at order $\alpha_s^4$
(like  corrections to the Bjorken (polarized)
DIS sum rule)  and {\em some} of
them---like the Adler function---even at next, in fact, five-loop, 
$\alpha_e \alpha_s^4$ order 
\cite{Baikov:2010je}. By
one-scale physical quantities we mean here scale invariant massless
correlators (or their proper combinations) in the Euclidean region\footnote{We do
not consider well known terms proportional to  various powers of $\pi$ which are routinely
generated during the procedure of analytical continuation to the Minkowskian
(negative) values of the momentum transfer $Q^2$.}.

The  appearance of $\zeta_4$ in a one-scale physical quantity has been 
demonstrated in \cite{Baikov:2017ujl} for the case
of the 5-loop  scalar correlator.


Very recently M.~Jamin and R.~Miravitllas have discovered that after
a transition to a new so-called C-scheme \cite{Boito:2016pwf} all terms
proportional to even zetas ($\zeta_4$ and $\zeta_6$ in the cases under
consideration) do disappear in the 5-loop scalar correlator as well as in
the 5-loop gluon correlator \cite{Jamin:2017mul} 
(both enter the hadronic decays of the Higgs boson 
\cite{Baikov:2005rw,Herzog:2017dtz}).
They also  suggested that the absence of even zetas after transition to the
C-scheme is an universal feature of {\em all} ${\cal O}(\alpha_s^5)$
physical quantities\footnote{
Please, pay attention  that the counting of  powers of $\als$  is definition depended, 
see  more details in the next Section.}
--- ``no-$\pi^2$ conjecture''.  

Later many more  
confirmations of the conjecture have been discussed in \cite{Davies:2017hyl,Chetyrkin:2017bjc,Ruijl:2018poj}.

The most interesting feature of the conjecture is a direct universal
connection between $\zeta_4$ contribution to physical quantities at order
$\alpha_s^5$ and the $\zeta_4$-term in the QCD $\beta$-function which appears
first at order $\alpha_s^5$
\cite{Baikov:2016tgj,Herzog:2017ohr,Luthe:2017ttg}.  To really appreciate the
mystery behind these cancellations we want to remind the reader of the
following simple facts:

\begin{enumerate}
\item   a bare physical  (massless!) quantity   depends on the bare coupling constant, 
$\alpha_s^B$;

\item   its renormalization is done  with the  replacement $\alpha_s^B = \mu^{2\ep}Z_a \alpha_s$;

\item   the charge  Renormalization Constant (RC)  $Z_a$   depends on the  five-loop coefficient in   the
$\beta$-function --- $\beta_5$ --- starting from  the {\em fifth}  order, $\alpha_s^5$; 

\item  as a result the renormalized physical quantity starts to ``feel''
  $\beta_5$ only at the  astonishingly large {\em sixth} order in $ \alpha_s$;

\item    for the case of the scalar correlator the contribution of order $\alpha_s^6$ corresponds to
the fabulously large  7-loop level.
\end{enumerate}


In the present work we suggest a natural explanation of  the mystery above
as well as a rigorous proof of the no-$\pi^2$ conjecture  (at least for the
5-loop level), by demonstrating that the $\zeta_4$ term in the
$\beta$-function is, in fact, not independent but must meet a simple
factorization formula: \beq
\beta_5^{\z_4} = \frac{9}{8}  \beta_1 \, 
 \beta_4^{\z_3}
{},
\label{z3z4:beta}
\eeq
where the $\beta$-function is defined as:
\beq
\mu^2 \frac{\mathrm d}{\mathrm d \mu^2} \, a(\mu) = a\, \beta(a) = a \sum_{ i \ge 1 } \beta_i \, a^i 
{}, \ \ a=\frac{\alpha_s}{4\,\pi} = \frac{g_s^2}{16\, \pi^2}
\EQN{beta:def}
{}
\eeq
and
the upper-script $\z_i$  means
\beq
 F ^{\z_i} = \lim_{\z_i \to 0} \frac{\prd}{\prd \z_i} F 
\ \ \mbox{and (for future reference)} \ \  
F ^{\z_i \z_j} = \lim_{\z_i \to 0} \frac{\prd}{\prd \z_i} F^{\z_j} 
{}.
\eeq

The validity of the factorization formula will be proven {\em without} any
loop calculation and (almost) without any information on the $\beta$-function.
The proof is based solely on the general considerations about the structure of the
4-loop massless propagators discussed in Section of \cite{Baikov:2010hf} and the fact that
the 3-loop $\beta$-function is free from any zetas \cite{Tarasov:1980au,Larin:1993tp}.
 
Note that factorization in \re{z3z4:beta} is not trivial even for  QCD with the  SU(3)  gauge group
(the coefficient $\beta_4$ was first computed in \cite{vanRitbergen:1997va} and confirmed in \cite{Czakon:2004bu})
:
\begin{eqnarray}
\phantom{\frac{\prd}{\prd \zeta_4} \,  \beta_5} 
&\phantom{=}& \phantom{\frac{9}{8}} \hspace{3em}  
\beta_1 \hspace{7em} (\prd/\prd\zeta_3) \beta_4
\nonumber \\
\frac{\prd}{\prd \zeta_4} \,  \beta_5 &=& \frac{9}{8} \Big(\frac{2}{3}\,n_f-11\Big)
\Big(
-\frac{6472}{81}\,n_f^2
+\frac{6508}{27}\,n_f
-3564\Big)
\nonumber
\end{eqnarray}
while for a general case it takes the form:
\begin{eqnarray}
\phantom{\frac{\prd}{\prd \zeta_4} \,  \beta_5} 
&\phantom{=}& \phantom{\frac{9}{8}} \hspace{3em}  
\beta_1 \hspace{7em} (\prd/\prd\zeta_3) \beta_4
\nonumber \\
\frac{\prd}{\prd \zeta_4} \,  \beta_5 
&=&
\frac{9}{8} \Big(
\frac{4}{3}\,n_f\,T_F
-\frac{11}{3}\,C_A
\Big)
\Big(
\frac{44}{9}\,C_A^4
-\frac{136}{3}\,C_A^3\,n_f\,T_F
\nonumber\\
&\phantom{=}&
+\frac{656}{9}\,C_A^2\,C_F\,n_f\,T_F
-\frac{224}{9}\,C_A^2\,n_f^2\,T_F^2
-\frac{352}{9}\,C_A\,C_F^2\,n_f\,T_F
\nonumber\\
&\phantom{=}&
-\frac{448}{9}\,C_A\,C_F\,n_f^2\,T_F^2
+\frac{704}{9}\,C_F^2\,n_f^2\,T_F^2
-\frac{704}{3}\,\frac{d_A^{abcd}\,d_A^{abcd}}{N_A}
\nonumber\\
&\phantom{=}&   
+\frac{1664}{3}\,\frac{d_F^{abcd}\,d_A^{abcd}}{N_A}\,n_f
-\frac{512}{3}\,\frac{d_F^{abcd}\,d_F^{abcd}}{N_A}\,n_f^2
\Big).
\end{eqnarray}

We will derive a variety of exact relations similar to (\ref{z3z4:beta}) valid
for arbitrary \MSbar\ \cite{'tHooft:1973mm,Bardeen:1978yd} anomalous
dimensions (ADs).  We also will find new identities which relate contributions
proportional to even zetas in (not necessarily physical) massless correlators
and terms proportional to odd zetas in the corresponding ADs.

In addition we suggest a simple proof of the no-$\pi$ theorem valid for any
number of loops and every massless correlator (including those with non-zero
AD) provided the participating Feynman integrals meet a
simple condition to be specified later.  The condition is shown to be
fulfilled for all currently known examples of the theorem.  We also construct
explicit expressions for the $\pi$-dependent contributions to  6-loop 
ADs and   $\beta$-function valid  for a generic
one-charge theory.

If not otherwise stated we will assume  the so-called $G$-scheme for
renormalization \cite{Chetyrkin:1980pr}. The scheme is natural for massless propagators.  All
ADs, $\beta$-functions and $Z$-factors are identical in
\MSbar- and G-schemes.  For (finite) renormalized functions there exists a
simple conversion rule. Namely, in order to switch from an  $G$-renormalized
quantity to the one in the \MSbar-scheme one should make the following
replacement in the former: $\ln{\mu^2} \to \ln{\mu^2} +2$ (
$\mu$  is the renormalization scale, the limit of $\ep \to 0$ is understood).

\section{Setup}
Let $F$ be any {\em p-function}, that is a  Green function or a 2-point correlator (or even some
combination  thereof), expressible in terms of massless propagator-like
Feynman integrals (to be named  p-integrals below). As an example one could
have in mind any of the 11 objects computed and renormalized at the 4-loop level
in \cite{Ruijl:2017eht}.  $F$ may not necessarily be a gauge invariant object, if
there is a gauge dependence we will always assume the Landau gauge fixing
condition below.  With such convention at hand we can and will effectively ignore the
gauge parameter dependence (if any) in all objects under consideration here.
The (renormalized)  p-function $F \equiv F_R$ can be  naturally represented   as
\bal
F_n(a,\lQ) & = 1  +\sum_{ 1\le i \le  n }^{0 \le j \le i} g_{i,j} \, \lQ^j \, a^i
\EQN{G:def}
{},
\end{align}
where $a=\frac{\alpha_s(\mu)}{4\,\pi}$, $\lQ= \ln \frac{\mu^2}{Q^2}$ and $Q$ is
an (Euclidean) external momentum. The integer $n$ stands for the (maximal)
power of $\alpha_s$ appearing in the Feynman diagrams contributing to
$F_n$. The $F$ without $n$ will stand as a shortcut for a formal series $F_\infty$.
In terms of bare quantities $F$ is written as
\beq
F = Z\, F_B(a_B,\lQ), \hspace{2cm} Z  = 1 +\sum_{ i \ge 1}^{1 \le j \le i} 
Z_{i,j} \, \frac{a^i}{\ep^j}
\EQN{G:GB}
{},
\eeq
with   the bare coupling  constant and the corresponding RC being
\beq
a_B = \mu^{2\ep} Z_a \,a , \hspace{2cm}  Z_a   = 1 +\sum_{  i \ge 1 }^{1 \le j \le i} 
\Bigl(Z_a\Bigr)_{i,j} \, \frac{a^i}{\ep^j}
{}.
\eeq
The evolution equation for F reads:
\beq
\Bigl(\frac{\prd}{\prd \lQ}\, + \be\,a\, \frac{\prd}{\prd a}\Bigr) F = \g\, F
\EQN{G:RG:evol}
{},
\eeq
with 
the AD 
\beq
\g(a) =  \sum_{ i \ge 1 } \g_i \, a^i, \ \    \g_i = -i Z_{i,1}  
\EQN{gama:def}
{}.
\eeq
The coefficients of the $\beta$-function $\beta_i$ are related to  $Z_a$ in the 
standard way:
\beq
\beta_i = i \left(Z_a\right)_{i,1} 
\EQN{beta:coef:def}
{}.
\eeq
Note that if the AD $\g$ happens to be equal $\beta$ for a
p-function $F$ (an explicit  example will be considered later) then the latter
should be renormalized with $Z\equiv \left(Z_a\right)^{-1}$ (due to different
signs in the above expressions for $\g_i$ and $\beta_i$).

Currently our ability to compute p-functions in QCD is  limited to 4 loops and
the corresponding ADs to 5 loops (see, e.g. \cite{Baikov:2015tea}).  
If $\g$ is not vanishing then there are  two ways to construct a
scale-invariant version of $F$.

Let us consider them in  turn.
The first one is to construct an object:
\bal
\hat{F}^{\, \mathrm si}(a,\lQ)_{n+1} &= (a)^{\frac{-\g_1}{\beta_1}} \exp \Biggl\{
-
\int_0^a \frac{\mathrm{d} x}{x} \Bigl[ 
\frac{\g(x)}{\beta(x)} - \frac{\g_1}{\beta_1 }
\Bigr]
\Biggr\}\, F_n(a,\lQ) 
\\
&=  (a)^{\frac{-\g_1}{\beta_1}}\,
\Bigl(
1 +\sum_{ 1\le i \le n }^{0 \le j \le i} \hat{d}_{i,j} \, a^i \, \lQ^j
\Bigr)
\EQN{SI:def}
{},
\end{align}
where $a=a(\mu)$. Note that $\hat{F}^{\, \mathrm si}_{n+1}$  is built from $F_n$ and
the (n+1)-loop AD $\g$.

\ice{
 it is understood  that only terms $a^i$ with $i \le $ 
$n$ are  kept  in the rhs of \re{SI:def} (the first  factor is not counted!).
Note that the maximal  order in $\alpha_s$ in \re{SI:def}   is the same as in the original   function
\re{G:def}. Later we consider separately  the case of  a correlator of two currents where the  loop number of 
$\hat{G}^{\mathrm si}_n$ is by one larger.
}

We will be interested in even zetas appearing in the coefficients $\hat{d}_i
\equiv \hat{d}_{i0}$\ice{(the coefficients $\hat{g}_{ij}$ with $j \ge 1$ will be
discussed later)}. It is a well-know fact (see discussion in 
Section \ref{sec:structure}) that even zeta may
appear only starting from 3 loops for correlators and from 4 loops for
ADs. In addition $\beta_4$ does not depend on $\zeta_4$
(this is a   peculiar property of $\beta_4$ in QCD).
   An account of this directly leads to the following
representations for the (potentially) $\pi$-dependent contributions
to $\hat{d}_3$ and   $\hat{d}_4$:
\bal
\hat{d}_3 & \epi -\frac{1}{3\beta_1}\Bigl( \g_4 - 3\,\beta_1 g_3 \Bigr) 
\EQN{hd3}
{},
\\
\hat{d}_4 & \epi -\frac{1}{4\beta_1}\Bigl( \g_5 -\frac{\g_1\,\beta_5}{\beta_1} -\frac{\g_4\,\beta_2}{\beta_1}
 + \frac{4\,\g_4 g_1}{3} - 4 \beta_1 g_4 
 \Bigr)
\EQN{hd4} 
{},
\end{align}
where the sign $\epi$ means equality modulo  $\pi$-independent terms.
Note that in eq. \re{hd4} we have used the 
fact (proven in \cite{Baikov:2010hf} and discussed later\footnote{
See eq. \re{identity:1}.}) that $\hat{d}_3  \epi 0$.
 
Unlike the 3-loop  level, the  4-loop  coefficient $\hat{d}_4$ does 
provide  no-zero contribution proportional to $\zeta_4$  in  every  available example with $\g \not = 0$, 
but   terms with $\zeta_6$ and $\z_3\, \z_4$  never show up. 
It is of interest that   the second combination which does appear in many master integrals
\cite{Baikov:2010hf,Lee:2011jt}  never contributes  to  
{\em all} known 4-loop  renormalized correlators
(not necessarily  scale-invariant)
 and 
5-loop ADs.  The reason for it will be clarified in Section \ref{sec:structure}.  

The remarkable  observation first   made in  \cite{Jamin:2017mul}  
can,  in our setup,  be 
precisely  formulated as follows:
\beq
\hat{d}_4  \epi   \frac{\g_1\, \beta_5 }{3\,\beta_1^2}
\EQN{discovery}
\eeq
or, in an explicit form, 
\beq
\hat{d}_4 -  \frac{\g_1\,\beta_5}{3\,\beta_1^2} \epi  -\frac{1}{4\beta_1}\Bigl(  
\frac{\g_1\,\beta_5}{3\,\beta_1} + \g_5  -\frac{\g_4\,\beta_2}{\beta_1}
 + \frac{4\,\g_4 g_1}{3} - 4 \beta_1 g_4 
 \Bigr)
\epi 0
\EQN{discovery2}
{}.
\eeq
We will prove \re{discovery2} in Section \ref{interplay:generic:z34}.

If we change   the renormalization scheme as follows:
\beq
a = \bar{a} \, ( 1 + c_1\, \bar{a} +c_2\,  \bar{a}^2 + c_3 \, \bar{a}^3 +  c_4 \bar{a}^4 )
\EQN{new:scheme:1}
{},
\eeq
with $c_1, c_2$ and  $c_3$ all free from even zetas and with  
\beq
c_4 \epi \frac{1}{3} \, \frac{\beta_5}{\beta_1} \epi \frac{3}{8}\,  \beta_4^{\z_3}\, \z_4
\EQN{new:scheme:2}
{},
\eeq
then the function $\hat{F}^{\, \mathrm si}_5(\bar{a},\lQ)$ as well the (5-loop) $\beta$-function
$\bar{\beta}(\bar{a})$ both loose any dependence on even zetas.
We will call  the class of renormalization  schemes for which
\beq
\bar{\beta}(\bar{a}) \epi {\cal O}(\bar{a}^{L+1}) \ \ \mbox{and} \ \ c_1 \epi 0
\EQN{beta:pi:ind}
\eeq
as  $\pi$-independent to L loops  schemes. 
Note that any  two $\pi$-independent (to L loops) schemes are related by transformation   \re{new:scheme:1}
with all coefficients $c_1 \dots c_{L-1}$  being $\pi$-free.

An example of such a  scheme has been  recently 
considered in \cite{Boito:2016pwf} under the name ``C-scheme''. 
In this scheme the $\beta$-function is    $\pi$-independent in all orders.
In particular,
\beq
c_4 = \frac{1}{3} \, \frac{\beta_5}{\beta_1}
\EQN{c:scheme}
{}.
\eeq

The second way to construct a scale-invariant object from $F$ is also
well-known\footnote{ In DIS similar objects are called "physical anomalous
  dimensions", see, e.g. \cite{Davies:2017hyl}.  To the best of our knowledge,
  the trick was first applied for constructing a $\pi$-free version of a
  2-point  correlator in \cite{Vermaseren:1997fq}.}:

\beq
F^{\mathrm si}_{n+1}(a,\lQ) = \frac{ \prd }{\prd \lQ} \left(\ln F\right)_{n+1}
\equiv \Biggl(\frac{\left( \g(a) - \beta(a) a \frac{\prd}{\prd a}\right)\,F_n}{F_n}\Biggr)_{n+1}
{}.
\EQN{dG1}
\eeq 
Note that $F^{\mathrm si}_{n+1}(a,\lQ)$ starts from the first power of the coupling
constant $a$ and is formally   composed from ${\cal O}(\alpha_s^{n+1})$ Feynman  diagrams.
In the same time  is can be completely restored from $F_n$ and the  $(n+1)$-loop AD $\g$.
This dual nature of $F^{\mathrm si}_{n+1}$ plays a vital  role in  all our
considerations in the next two Sections.

It is convenient to write
\beq
F^{\mathrm si}_{n+1}(a,\lQ) = a\,\left(
\g_1 + \sum_{ 1\le i \le  n }^{0 \le j \le i} d_{i,j} \, \lQ^j \, a^i
\right)
\EQN{dG2}
{}.
\eeq
The coefficients $d_i = d_{i0}$ meet, obviously, the  equations  (parallel to  
\re{hd3} and \re{hd4}):
\bal
{d}_3 & \epi  \g_4 - 3\,\beta_1 g_3 
{},
\EQN{d3}
\\
{d}_4 & \epi  \g_5 -\frac{\g_4\,\beta_2}{\beta_1}
+ \frac{4\,\g_4 g_1}{3} - 4 \beta_1 g_4
{}.
\EQN{d4} 
\end{align}
\ice{
G5siz4a5 = Coef[G5siz4, a^5] // InputForm
a5 - (a4*b2)/b1 + (4*a4*g1)/3 - 4*b1*g4
    }
Note, that up to a common factor $\hat{d_3}$ is equal to  $d_3$; the same is true 
for the  pair $\hat{d_4}$ and $d_4$ except for a term proportional to $\beta_5$ which
{\em does not appear at all}  in $d_n$ if $n \le 4$. A transition to a  $\pi$-independent
renormalization scheme  according to  \re{new:scheme:1} and \re{new:scheme:2}  leads to
\beq
F^{\mathrm si}_5(\bar{a},\lQ) = a\,\left(
\g_1 + \sum_{ 1\le i \le  4 }^{0 \le j \le i} \bar{d}_{i,j} \, \lQ^j \, \bar{a}^i
\right)
\EQN{barG5}
{},
\eeq
with 
\beq
\bar{d}_2 \epi d_2,\  \bar{d}_3 \epi d_3, \
 \bar{d}_4 \epi d_4 +  \frac{\g_1 \beta_5}{3\,\beta_1} \epi
 \frac{\g_1\,\beta_5}{3\,\beta_1} +\g_5 -\frac{\g_4\,\beta_2}{\beta_1}
 + \frac{4\,\g_4 g_1}{3} - 4 \beta_1 g_4 
\EQN{new:scheme:3}
{}.
\eeq

Eqs.  \re{discovery2} and \re{new:scheme:3}  explicitly demonstrate that if, for  a given p-function $F_4$, 
its scale invariant version $F^{\mathrm si}_5(a,\lQ)$  is $\pi$-free in  a $\pi$-independent 
renormalization scheme then the function   $\hat{F}^{\mathrm si}_5(a,\lQ)$ is also $\pi$-free, and vice versa.

Our main aim is to   really understand and, consequently, to  prove the equality (which happens to be true
in all known so far examples)
\beq 
d_4 + \frac{1}{3} \, \frac{\g_1\,\beta_5}{\beta_1} \epi 0
\EQN{discovery:3}
{}.
\eeq
The equality is  naturally separated into two pieces
\begin{align}
d_4 + \frac{1}{3} \, \frac{\g_1\beta_5^{\z_4}}{\beta_1}\, \z_4 &\epifour 0
{},
\\
d_4   &\episix 0
{}.
\end{align}

Assuming  the factorization formula \re{z3z4:beta} (which will be derived  in Section \ref{interplay:special}),
we will prove in what follows that 
\begin{align}
d_4 + \frac{3}{8}\, \g_1\,\beta_4^{\z_3} \, \z_4&\epifour 0
{},
\\
d_4   &\episix 0
{}.
\end{align}


\section{\EQN{interplay:generic:z34} Interplay between $\z_3$, $\z_4$: generic case}

In general representation \re{dG1} is very suitable to discuss the interplay
of various transcendental contributions to $F$, $\beta$ and $\g$.  This is
because, unlike \re{SI:def}, $F^{\mathrm si}_n(a,\lQ)$ is a combination of a 
finite number of Feynman diagrams with a very simple renormalization mode:
$F^{\mathrm si}(a,\lQ) = F^{\mathrm si}_B(a_B =Z_a\, \mu^{2\ep}a,\lQ)$  (as we discussed
already in the Introduction). 

The {(bare)} function $F^{\mathrm si}_{5,B}$  is composed  from
1,2,3,4 and 5-loop bare p-integrals\footnote{In some cases (correlators
of composite operators) also 6-loop p-integrals might  contribute 
(see a discussion in Section
\ref{pi-safe:5and6L}).}.
We will call a (bare) $L$-loop p-integral $F(Q^2,\ep)$ {\em $\pi$-safe} 
if the $\pi$-dependence of its pole in $\ep$ {\em and} constant part
can be completely absorbed  into the properly  defined ``hatted'' odd zetas.
\ice{
if its
pole in $\ep$ {\em and} constant parts can be rewritten solely in terms of
properly defined ``hatted'' odd zetas\ice{\footnote{ By odd zetas we mean  here  any
transcendental objects which vanish identically after formal setting $\pi=0$ .}}.
}
The first observation of a  non-trivial class  of $\pi$-safe  p-integrals --- all
3-loop ones --- was made in   \cite{Broadhurst:1999xk}.
An extension of the observation on the  class of all 4-loop p-integrals 
was performed in \cite{Baikov:2010hf}.
Here it was shown that, given an arbitrary 4-loop p-integral,    
its pole in $\ep$ {\em and} constant part 
depend on even zetas {\em only} via
the following combinations:
\ice{
\footnote{It is clear that one can not really  hide {\em all} 
$\pi$-dependence inside $\hat{\zeta}_3$ and $\hat{\zeta}_4$; it will  reappear at higher orders
of $\ep$-expansion which, however,   never contribute to renormalized $D=4$ four-loop p-integrals.
See an elaborated discussion in Section ?}.
}
\beq
\hat{\zeta}_3 \defas \zeta_3+\frac{3 \ep}{2}{\zeta_4}-\frac{5 \ep^3}{2}{\zeta_6},
\,\, 
\hat{\zeta}_5\defas\zeta_5+\frac{5\ep}{2}\zeta_6 \ \ \  \mbox{and} \ \  \ \hat{\zeta_7} \defas \z_7.
\label{hat:zetas}
\eeq 
Note that the   $\pi$-safeness of, say,  all $L$-loop p-integrals ensures 
that for any  $L'$-loop $p$-integral  $F$  with $L' < L$ the $L$-loop combination 
$\frac{1}{\ep^{L-L'}} F $  is a $\pi$-safe one  with  the same  choice of the 
hatted zetas.

We will proceed, assuming for the moment that all 
p-integrals contributing to $F^{\si}_5$ are $\pi$-safe.
The assumption is discussed and eventually
proved in Section \ref{pi-safe:5and6L} (at least for the particular classes of p-integrals which we
encounter in the present paper).

Any renormalized  p-function  $F_{5,R}$
can, by definition, be   presented as
\beq
F_{5,R} = F_{5,B} + \sum_{i} Z_i\,  p_i
\EQN{G5R}
{},
\eeq
where $Z_i$ are some UV counterterms and $p_i$ are p-integrals.  Let us assume
for a moment that all $Z_i$ in  \re{G5R} are $\pi$-free
and rewrite all (including those contributing to $F_n^B$) p-integrals
in \re{G5R} in terms of hatted zetas. 
Then the rhs of \re{G5R} will depend on
hatted odd zetas only (in the limit of $\ep \to 0$).  The finiteness of
$F_{5,R}$ would ensure then the absence of any $\pi$-dependent terms in it in the $\ep\to 0$
limit.  Clearly, if some of $Z_i$ do contain odd and/or even  zetas then 
there should exist  some constraints relating contributions with odd and even zetas
to the p-function  and  depending on a  precise pattern of hiding $\pi$-dependent 
terms inside hatted odd zetas.  These constraints will be the main tool in our
considerations.

Let us consider first $F^{\mathrm si}_4$. It is renormalized with $Z_a$ in  three  loops
which is free from any zetas for QCD.  As a result,  $F^{\mathrm si}_4$ should be free from even zetas which means the fulfillment of the identity
\beq
d_3\epi \g_4 -3\,\beta_1 g_3 \epi 0 
\EQN{identity:1}
{}.
\eeq

\newcommand{\LgR}{\left( \g \right)}

\newcommand{\LZaR}{\left( Z_a \right)}

\newcommand{\LZR}{\left( Z \right)}

Now let us turn to  the  renormalization of the very function $F_3$. Its
renormalization mode includes  $Z_a$ (in 2 loops, free from any zetas) and
$Z_\g$ (in 3 loops).  Note that $(Z_\g)_3$  depends generically on the
combination $\z_3/\ep$.  By definition of the AD
\beq
(Z_\g)_{3,1}^{\z_3}  = -  \frac{1}{3} \g_3^{\z_3}
{}.
\eeq

Let us write the renormalized function $F_3$ as follows\footnote{
For simplicity we set $\mu=1$ in all  formulas below.} 
\beq
F_3 = (Z_\g)_{3,1}^{\z_3} \,\z_3 \, \frac{a^3}{\ep} + \sum_i  z_i(\ep)\, p_i\, a^3 + 
\ \ \mbox{terms of order $a^2$ and lower} 
\EQN{G3:finite}
{},
\eeq
where ${p_i}$ are some bare  p-integrals  and  the coefficients ${z_i}$ are  
some polynomials in $1/\ep$ 
(the  constant term is allowed!) with  {\em rational} coefficients.
Assume now  that  the p-integrals in the  above sum are expressed via hatted 
zetas. The coefficient in front of $\hat{\zeta}_3$ should  then be equal 
\[
-(Z_\g)_{3,1}^{\z_3} \frac{a^3}{\ep} \equiv  \g_3^{\z_3}\,\frac{a^3}{3\, \ep}
{}.
\]
Expressing  back $\hat{\z_3}$ via normal zetas  we arrive at the identity:
\beq
g_3^{\z_4} \equiv  \frac{1}{2} \g_3^{\z_3}
{}.
\eeq
Now a look on   \re{identity:1}   immediately leads 
to  relations valid for any  AD $\g$ 
\beq
\g_4^{\z_4} = \,\frac{3}{2} \,\beta_1 \g^{\z_3}_3
\EQN{identity:2}
{}.
\eeq

The fact that $\beta_3 \epi 0$ is a particular feature of QCD.  
The above treatment  can  be easily
extended for a general  case of  $\beta_3^{\z_3} \not= 0$. The corresponding 
generalizations of eqs.~\re{identity:1} and  \re{identity:2} read
\beq
d_3\epi \g_4 -3\,\beta_1 g_3 \epi -\frac{1}{2} \beta_3^{\z_3}\, \g_1 \, \z_4
\EQN{identity:1:extended}
{},
\eeq
\beq
\g_4^{\z_4} = \,\frac{3}{2} \,\beta_1 \g^{\z_3}_3 -  \frac{1}{2} \beta_3^{\z_3}\,\g_1 
\EQN{identity:2:extended}
{}.
\eeq
For  a particular case of $\g \equiv \beta$ we arrive at
\beq
\beta_4^{\z_4} = \beta_1 \beta^{\z_3}_3 
\EQN{identity:2:beta}
{}.
\eeq

\section{\EQN{interplay:generic:z56} Interplay between  $\z_5$, $\z_6$: QCD case}

Let us now add one more loop and consider $F_5^{\si}$. It is  renormalized
with the  RC $Z_a$ (taken in the  4-loop approximation):
\beq
F_5^{\si} =  \g_1\,(Z_a)_{4,1}^{\z_3} \left(\hat{\z}_3 -\frac{3\ep}{2}\z_4 \right)
 \frac{a^5}{\ep} +
 \unl{\sum_{i,1 \le j \le 5}  z_{i,j}(\ep)\, \hat{p}_{i,j}\, a^j} +{\cal O}(\ep^0)
\EQN{G5si:z5}
{},
\eeq
where $\hat{p}_{i,j}$ stands  for $p$-integrals expressed in  terms of hatted (odd!) zetas
and  $z_{i,j}(\ep)$ are completely rational polynomials in $1/\ep$.
The  finiteness of   \re{G5si:z5}  means that  the  term   proportional to  $\hat{\z}_5$
in the underlined sum in \re{G5si:z5} should  have a  finite coefficient. Thus, 
the whole p-function
$F_5^{\si}(a,\lQ)$ should be  free from $\zeta_6$ in the limit $\ep \to 0$. This could be  precisely  written as
$d_4 \episix 0$ or, equivalently,
\beq
  \g_5 -\frac{\g_4\,\beta_2}{\beta_1}
+ \frac{4\,\g_4 g_1}{3} - 4 \beta_1 g_4 \episix \g_5- 4 \beta_1 g_4  \episix 0
\EQN{d4:epi6}
{},
\eeq
where  we have used  the  fact that  $\g_4    \episix 0 $ (see Table \ref{tab1}).
On the other hand, the pole part of the coefficient in front of $\hat{\z}_3$
in the underlined sum must be equal to  $-\g_1\,(Z_a)_{4,1}^{\z_3}/\ep $. This
directly leads to the identity
\beq
d_4 \epi -\g_1\,\frac{3}{8} \beta^{\z_3}_4  \z_4 
\EQN{noPi:5l}
{}.
\eeq

Thus, we have  proved the  no-$\pi$ theorem: for every  (QCD!) p-function  
$F_4$  the corresponding 
scheme-invariant function $F_5^{\si}$ is $\pi$-independent. 
We will rederive it for   generic functions $F_{n+1}^{\si}$ it  more systematic way in Section \ref{Big:no:pi}.

Let us write now the renormalized version of $F_4$ in a form explicitly showing 
all possible 
zetas  coming from RCs.  The form reads
(an upper index $n\ell$ in $\hat{p}_i^{n\ell}$ means that the  p-integral $p_i$ 
has $n$ loops)
\beq
F_4 \episix  a^4\,\Bigl[ Z_{3,1}^{\z_3}\,\frac{\z_3}{\ep} \sum_i \hat{p}_i^{1\ell} 
+
 Z_{4,2}^{\z_3}\,\frac{\z_3}{\ep^2}  +   Z_{4,1}^{\z_3}\,\frac{\z_3}{\ep}
+
 Z_{4,1}^{\z_4}\,\frac{\z_4}{\ep} +  Z_{4,1}^{\z_5}\,\frac{\z_5}{\ep} 
+\sum_{i} \hat{p}_i^{4\ell}\Bigr]
\EQN{G4:ren:z6}
{},
\eeq
where we have discarded all terms obviously  independent of  $\z_6$.
The finiteness of coefficient in front of $\z_5$ in \re{G4:ren:z6}
leads to the  equality
\beq
\left(
\sum_{i} \hat{p}_i^{4\ell}
\right)^{\hat{\z_5}} = - Z_{4,1}^{\z_5}/{\ep}
{}.
\eeq
Thus,
\beq
g_4 \episix -\frac{5}{2} Z_{4,1}^{\z_5}\,\z_6 \episix \frac{5}{8} \g_4^{\z_5}\,\z_6 
\eeq
and, finally, after  the  use of \re{d4:epi6}
\beq
\g_5^{\z_6} = \frac{5}{2} \g_4^{\z_5}\,\beta_1
{}.
\eeq

All derivations of this Section have been done for the QCD case, that is
assuming $\beta_3 \epi 0$ and $\beta^{\z_5}_4 = 0$. A generalization to a
generic one-charge theory can be also done along the same lines. But we postpone
this  to Sections \ref{Big:no:pi}, \ref{6loops} and \ref{6loops:2} where we will
suggest an   universal treatment.

\section{\EQN{interplay:special} The connection between $\z_3$ and $\z_4$ terms in the
QCD $\beta$-function}

As is well-known the use of the Landau  gauge    implies that 
\beq
Z_a = (Z^{\mathrm{gh}})^{-2}\, (Z^{\mathrm{gl}})^{-1} \ \ \mbox{and} \  \ \beta = -2 \g^{\mathrm{gh}}  - \g^{\mathrm{gl}}
{}, 
\eeq
where $\g^{\mathrm{gh}}$ and $\g^{\mathrm{gl}}$ stand for the ghost and gluon wave function  ADs respectively
\cite{Taylor:1971ff,Blasi:1990xz}.

Let us  consider the following  combination of the ghost and  gluon self-energies
\beq
F(a,\lQ) = (1 +\Pi^{\mathrm{gh}})^2\, (1 +\Pi^{\mathrm{gl}}) = (Z_a)^{-1}\,
 (1 +\Pi_B^{\mathrm{gh}})^2\, (1 +\Pi^{\mathrm{gl}}_B)
\EQN{G:beta}
{}.
\eeq
This  is a p-function with the evolution  equation  
\beq
\Bigl(\frac{\prd}{\prd L}\, + \be\,a\, \frac{\prd}{\prd a}\Bigr) F = \beta\, F
\EQN{G:RG:evol:2}
{}.
\eeq
The $\beta$-function coefficient   $\beta_3$ is  a rational number   while $\beta_4$ does 
depend on $\zeta_3$. This is  the  only specific information on the  $\beta$-function which is  needed 
for  our analysis of  the transcendental structure of $F_4$ and the coefficients $\beta_4$ and $\beta_5$ 
 below.  Indeed, the following is true:
\vspace{3mm}

\noindent
{\bf Lemma}
If $F$ is defined by eq.~\re{G:beta}   and $\beta_3^{\z_3} = \beta_4^{\z_5} = 0$ then
\\
\begin{enumerate}
\item
The coefficients $g_i \epi 0$ if $i \le 3$. 

\item
The coefficient $\beta_4   \epi 0$.

\item 
The coefficients $\left(Z_a\right)_{4,j}$  are rational  for $j \ge 2$.

\item The coefficient $g_4 \epifour \frac{3}{8}\, \beta^{\z_3}_4\,\z_4$.

\item 
The coefficient $\beta^{\z_6}_5 = 0$.
\\
The transcendental contributions to $\beta_4$ and $\beta_5$  are constrained by the following relation:
\beq
\beta_5^{\z_4} = \frac{9}{8}  \beta_1 \,  \beta_4^{\z_3}
\EQN{beta5:beta4:constraints}
{}.
\eeq

\end{enumerate}

\vspace{3mm}

\noindent
{\bf Proof}

\begin{enumerate}
\item  Indeed, the p-function $F_3$ is renormalized with the 3-loop RC $Z_a$  and, 
thus, can not depend on even zetas 
(see the discussion after \re{G5R} and Section 7.1 of \cite{Baikov:2010hf}).

\item  This is a direct consequence of the fact that   $\beta_4 \equiv \g_4$ and $\g_4 \epi 3\, \beta_1 \, g_3$.     

\item This is an obvious consequence of the well-known fact that higher poles
 in a RC $Z$  at loop order $(L+1)$ are completely fixed by  
 simple pole terms of $Z$ and $Z_a$
 with loop order not exceeding $L$ (so-called 't~Hooft's constraints
\cite{'tHooft:1973mm}). The constraints  read 
\beq
a\,\frac{\prd Z_{*,k+1}}{\prd a} =
\left[\beta\,  a \frac{\prd }{\prd a} - \gamma 
\right] 
 Z_{*,k}
{},
\EQN{Hooft:cond}
\eeq
with $Z_{*,k} = \sum_{i \ge k}  Z_{ik} \, a^i$.  
\item  The renormalized function $F_4$ can be  presented as: 
\beq
F_4 = -(Z_a)_{4,1}^{\z_3} \, \z_3\,\frac{a^4}{\ep} + \sum_i  z_i(\ep)\, p_i\, a^4
\ \  +\mbox{terms of order $a^3$ and lower} 
\EQN{G4:finite}
{}.
\eeq
Literally repeating considerations after eq.~\re{G3:finite} we arrive at:
\beq
g_4 \epi (Z_a)_{4,1}^{\z_3} \frac{3}{2}\, \z_4 \equiv \frac{3}{8} \beta_4^{\z_3}\, \z_4.
\EQN{Gsi:4}
\eeq

\item Now we consider the p-function $F^\si_{5B}$. By  construction, its renormalized version 
can be written as 
\beq
F^\si_5(a,\lQ) = F^\si_{5,B} (a_B,\lQ), \ \ \ a_B = \mu^{2\ep} (Z_a)_4\, a
{},
\EQN{Gsi:5:1}
\eeq
where we have  explicitly indicated that higher than 4-loop contributions to RC  $Z_a$ 
have no effect on renormalization of  $F^\si_{5,B}$ (as the latter starts from   $\beta_1\,a_{{}_B}$). 
As a result we have a representation 
\beq
F_5^\si = (Z_a)_{4,1}^{\z_3} \, \z_3\, \frac{a^5}{\ep} \beta_1   + \sum_i  z_i(\ep)\, p_i\, a^5 
\ \ +\mbox{terms of order $a^4$ and lower} 
\EQN{Gsi:5:2}
{},
\eeq
where all $z_i$ are polynomial in $1/\ep$ with {\em rational} coefficients and $p_i$ are some 
p-integrals with loop number  less or  equal  5.
Assuming that all p-integrals in \re{Gsi:5:2} are rewritten in term of hatted zetas  and  requiring  finiteness 
of $F_5^\si$ at $\ep\to 0$   we immediately obtain:
\beq
d_4 \epi -\frac{3}{8} \beta_1\,\beta_4^{\z_3}\, \z_4   
\EQN{Gsi:5:3}
{},
\eeq
where we  have  used eq. \re{hat:zetas} and the identity $(Z_a)_{4,1} = \frac{1}{4}\beta_4$.
On the other  hand,  from \re{d4} and first two statements of the Lemma  we have 
\beq
d_4 \epi \beta_5   -\frac{\beta_4\,\beta_2}{\beta_1}
+ \frac{4\,\beta_4 g_1}{3} - 4 \beta_1 g_4 \epi \beta_5  - 4 \beta_1 g_4
\EQN{Gsi:5:4}
{}.
\eeq
Finally, by combining eqs. \re{Gsi:4}, \re{Gsi:5:3} and \re{Gsi:5:4} we arrive
at \re{beta5:beta4:constraints}.
\end{enumerate}

Thus,  the  puzzle discussed in the Introduction is solved:  the surviving $\z_4$  in (generic) p-functions 
$\hat{F}^{\, \mathrm si}_5$ and in 
${F}^{\, \mathrm si}_5$ are indeed  removed  by the   $\z_4$ piece of the 5-loop coefficient $\beta_5$
which, however, is generated  (and completely  fixed) by  $\beta_1$ and the 
$\z_3$  term in the  4-loop $\beta$-function.  
\vspace{2cm}


\renewcommand{\xcancel}[1]{{}}

\section{\label{sec:structure} The $\pi-\zeta-\ep$ structure of p-integrals and renormalization
constants}

\begin{center}
\begin{table}
\begin{center}
\begin{tabular}{ |c|c||c|c| } 
\hline
 L  & p-integrals  & L+1 &  Z 
\\
\hline
0  &  rational                          &  1  & $\mbox{rational}/\ep$ \\
1  &  rational/$\ep$                          &  2  & $\mbox{rational}/\ep^2$ \\
2  & $\z_3$                             &  3  &   $\z_3/\ep$ \\
3  & $\z_3/\ep,\, \z_4,\, \z_5$                 &  4  &  $\z_3/\ep^2,\, \{\z_4, \z_5\}/\ep$ \\
4  & $\z_3/\ep^2,\, \{\z_4, \z_5\}/\ep,\, \z_3^2,\, \z_6,\, \z_7,\, \xcancel{\z_3\z_4}$   &  5  %
&   $\z_3/\ep^3,\, \{\z_4, \z_5\}/\ep^2,\, \{\z_3^2, \z_6, \z_7, \xcancel{\z_3\z_4}\}/\ep$  \\
\hline
\end{tabular}
\caption{
\EQN{tab1}
\small The structure of p-integrals  (expanded in $\ep$ up to and including the  constant $\ep^0$ part)
and  RCs  in dependence on the loop number $L$. The 
inverse power of $\ep$ stands for the {\em  maximal} one  in generic case; in particular cases it
might be less. 
}
\end{center}
\end{table}
\end{center}
It is instructive to look into the structure of  the results obtained in 
\cite{Ruijl:2017eht,Chetyrkin:2017bjc} for
11 4-loop QCD p-functions and corresponding 5-loop ADs  as displayed by  Table \ref{tab1}.
Is it possible to understand the regularities shown in  Table \ref{tab1}? Sure.
\vspace{5mm}

\begin{enumerate}

\item
The rationality\footnote{In the sense of absence any irrational constants.}
of any 1-loop p-function is essentially made by hand by the choice of the
G-scheme.  
1-loop RC are, in fact, rational in any
minimal scheme be it MS-, \MSbar- or G-scheme.  The irrational
structure of the p-functions at 2,3 and 4 loop level directly follows from
explicit results for the corresponding master integrals as found in \cite{Baikov:2010hf}.

\noindent
The  fact that the set of zetas appearing in ($L+1)$ RCs is the same as the one
displayed by $L$-loop $p$-integrals directly follows from the following theorem 
proven in \cite{Chetyrkin:1984xa}.

\vglue 0.2cm
{\bf Theorem 1.} {\it Any (L+1)-loop UV counterterm for any
Feynman integral may be expressed in terms of pole and finite parts
of some appropriately  constructed $L$-loop  $p$-integrals.}
\vglue 0.1cm

The maximal power of $1/\ep$ with which a given zeta may appear in a $L$-loop
contribution to a renormalization constant Z directly comes from the 't Hooft
constraint \re{Hooft:cond}.

Note that points 1. and  2. are also covered by  remarks 1. and 2. on page 13 of \cite{Georgoudis:2018olj}.

 \item An attentive reader has probably  noticed that Table  \ref{tab1} does not 
   include a particular  combination of the transcendental weight 7, namely,  $\z_3 \z_4$.
  There is a  nice  explanation of the fact. 
  Indeed, let us take  an arbitrary 4-loop p-integral $F$. If we consider it as
  a function of the irrational constants $\z_3,\dots, \z_7$, then the
  following identity holds (see eq.~\re{hat:zetas}) 
\beq 
F(\z_3, \z_4, \z_5,  \z_6, \z_7) = 
{F}(\hat{\z_3},0, 
\hat{\z_5}, 0, 
\hat{\z_7}) + {\cal O}(\ep)
    {}.  
\eeq 
If  a particular combination $\z_3\, \z_4$  could appear in
    the finite part of the function 
\[F(\z_3, \z_4, \z_5, \z_6, \z_7)\]
then its
    hatted version would,  obviously,  contain a term proportional to
    $\hat{\z_3}^2/\ep$.  This would imply that the constant part of a 3-loop
    p-integral contained $\z_3^2$ which is in contradiction to the previous
    point.
\ice{
In order to a particular combination $\z_3\, \z_4$ may appear in
    the finite part of the function $F(\z_3, \z_4, \z_5, \z_6, \z_7)$ its
    hatted version should, obviously, contain a term proportional to
    $\hat{\z_3}^2/\ep$.  This would imply that the constant part of a 3-loop
    p-integral contained $\z_3^2$ which is in  contradiction to the previous
    point.
}

 
\end{enumerate}

\section{ \label{pi-safe:5and6L}$\pi$-free representation for 5- and 6-loop p-integrals} 

In  Sections \ref{interplay:generic:z34},\ref{interplay:generic:z56},  and
\ref{interplay:special} we have assumed that 
all encountered p-integrals are $\pi$-safe. 
In fact, for the treatment of these 3 Sections to be valid 
the proven existence of $\pi$-free
form for 4-loop p-integrals is just enough. Indeed, all currently known
$p$-functions $F_4$ can be separated into 2 classes.
\vspace{4mm}

\noindent

I. The lowest order contribution---the 1 in \re{G:def}---corresponds to a
tree-level (that is loopless) diagram. The class includes QCD propagators and
vertex functions (the  latter in the propagator-like  kinematical regime).  Here the only contributions to
the scale-invariant p-function $F_5^{\si}$ with loop number equal to 5 could
come from the derivative $\frac{ \prd }{\prd \lQ} F_5$. This derivative
annihilates the constant part of the corresponding 5-loop p-integrals and,
thus, does depend only on their {\em pole} parts.

 The Adler function in 5-loops also belongs to class I in spite of the fact
 that its lowest order term corresponds to a 1-loop diagram. This is because
 by construction (i) the function is scale-invariant and (ii) it is completely
 determined by the non-constant part of the polarization operator.

Thus, the no-$\pi$ theorem for class I requires the existence of the
$\pi$-free form only for the pole part of an arbitrary 5-loop p-integral
(Property I).

\vspace{4mm}

\noindent
II. 
Any correlator $\Pi(a,Q^2)$ of 2 composite operators with non-zero ADs. Here
the starting object $F(a,Q^2)$ is constructed by a repeated application of the
operator $\frac{ \prd }{\prd Q^2}$ to $\Pi$ in order to arrive to the
superficially convergent p-function $F$.  During this procedure the bare
p-integrals are effectively multiplied by $\ep$.  The lowest order
contribution to $F$ starts (at least) from the one-loop term. Current
technology allow to compute $F_4$ and the corresponding AD\footnote{which is,
  obviously, equal to the sum of the ADs of the both composite operators.},
$\g$, up to and including 5 loops.  As a result the p-function $F_5^{si}$ is
contributed by 6--loop diagrams! However, a little meditation shows that only
$1/\ep^2$ and higher poles in these integrals could really contribute to
$F_5^{si}$.

Thus, the  no-$\pi$ theorem for class II requires the existence  of the $\pi$-free form only for  
arbitrary 6-loop  p-integrals  multiplied by $\ep^2$ (Property II).

Both Properties I and II are direct  consequences of the following 

\noindent 
{\bf Theorem 2}.
{\it

\noindent 
Let us assume that a  $\pi$-free form exists for all $L$-loop p-integrals. Then 

\noindent 
1. any $(L+1)$-loop p-integral times $\ep$ is $\pi$-safe.

\noindent 
2. any $(L+2)$-loop p-integral times $\ep^2$ is $\pi$-safe.  
\vspace{3mm}
}
\\
\noindent 
{\bf Proof}.
\newcommand{\G}{\Gamma}

\newcommand{\La}{\langle}
\newcommand{\Ra}{\rangle}

1. Let us define the  \Ghat-scheme  by pretending that  hatted zetas  do not depend on $\ep$.
This means  that all p-integrals are assumed to be expressed  in term of the
hatted  zetas and that the extraction  of the  pole  part of a p-integral  is defined as:
\beq
\hat{K}  \Bigl( 
{\cal P}(\ep)\prod_j \hat{\zeta}_j \Bigr) \defas   \left(   \sum_{i < 0} {\cal P}_i\, \ep^j  \right)   \prod_j\hat{\zeta}_j  
\EQN{hatK}
{},
\eeq
with ${\cal P}(\ep) = \sum_i \ep^i {\cal P}_i$ being  a   polynomial in $\ep$
with rational coefficients.
The corresponding coupling constant  will be denoted as $\ha$.

Theorem~1 insures then  that  all 3,4, $\dots, (L+1)$ counterterms are 
$\pi$-free\footnote{We always assume that  the dependence of hatted zetas on normal ones is 
polynomial in $\ep$.} in the new  scheme. 
Let now $\La\G\Ra$ be an arbitrary $(L+1)$-loop p-integral (corresponding  to  a \mbox{Feynman graph $\G$}).  
Its \Ghat-renormalized (and,
consequently, finite at $\ep \to 0$) version reads 
\beq
R \,\langle\Gamma\rangle(Q^2,\mu^2) = \langle\Gamma\rangle(Q^2,\mu^2) + Z_\Gamma +
\fbox{$
\sum_{\gamma}Z_\gamma \langle\Gamma/\gamma\rangle(Q^2)
{} + \dots$}
\label{R1}
\eeq
Here $Z_\gamma$ is the  UV Z-factor corresponding to a OPI subgraph
$\gamma$ of $\Gamma$, $Z_\Gamma$ is the UV counterterm for the very 
Feynman integral  $\langle\Gamma\rangle$   and dots stand for contributions with two and
more UV subtractions.   
The finiteness of $R \,\langle\Gamma\rangle$ implies that
\beq
\ep \, \langle\Gamma\rangle(Q^2,\mu^2)  = -\ep\left(  Z_\Gamma +
\sum_{\gamma}Z_\gamma \langle\Gamma/\gamma\rangle(Q^2)
{} + \dots 
\right)        + {\cal O}(\ep)
{}.
\label{R2}
\eeq
Note that the rhs of eq.~\re{R1} includes  only bare p-integrals with loop
number not exceeding $L$  as well as $\pi$-free UV  counterterms.
Point 1. is   proven.

2. Let us  consider \re{R1} with $\La \G\Ra $ being 
an ($L+2)$-loop p-integral.
Every particular term in the boxed part of eq.~(\ref{R1}) is a product of some
($\mu$-independent!) Z-factors and a reduced Feynman integral, the latter by construction
includes a factor $(\mu^2)^{n \ep}$, with $n$ being its loop number.
The very $p$-integral $\langle\Gamma\rangle(Q^2,\mu^2)$ depends on $\mu$  via 
a factor $(\mu^2)^{(L+2)\,\ep}$.
Thus, after  applying the operator $\mu^2 \frac{\prd}{\prd\mu^2}$ to eq.~\re{R2}
we obtain
\beq
\ep^2 \, \langle\Gamma\rangle(Q^2,\mu^2)  = -\ep^2 \left( 
\sum_{\gamma}Z_\gamma \, 
\left( 
1 - \frac{L(\g)}{L+2}
\right)
\langle\Gamma/\gamma\rangle(Q^2)
{} + \dots 
\right)        + {\cal O}(\ep)
{},
\label{R3}
\eeq
where $L(\g)$ is the loop number of the subgraph $\g$. 
Let us demonstrate that eq.~\re{R3} indeed proves point 2.  First, in the
\Ghat -scheme every RC  $Z_\g$ in its rhs is  $\pi$-free as its  loop number can
not exceed $(L+1)$. Second, one can also check that every term 
$Z_\gamma \,\ep^2\, \La \Gamma/\gamma\rangle$ is $\pi$-safe p-integral. For instance, 
if $L(\g) = 1 $ then $\La \G/\g\Ra $ is an $(L+1)$-loop p-integral 
and $Z_\g = C\,/\ep$,  with $C$ being an $\ep$-independent constant. As a result
we find that 
\beq
Z_\gamma \,\ep^2\, \La \Gamma/\gamma\Ra = C \, \ep\,  \La \Gamma/\gamma\Ra 
\eeq
is a $\pi$-safe combination due to point 1.

\section{\EQN{Das5} Adler function  at order $\alpha_s^5$}

It has been conjectured in \cite{Jamin:2017mul} that  
{\it ``a $\z_4$ term should arise in the Adler function at order
$\alpha_s^5$  in the \mbox{\rm \MSbar}-scheme, and that this term is expected to disappear in the C-scheme as well''}.

Our analysis certainly confirms the conjecture and upgrade it to a firm
prediction {\bf provided} that the product $\ep\,F(Q^2)$ is $\pi$-safe for
every 6-loop p-integral $F(Q^2)$.  Due to \mbox{Theorem 2} the condition is
equivalent to the $\pi$-safeness of the whole class of 5-loop p-integrals.
Fortunately, very recently  the last statement has received  strong 
 support from  explicit evaluation of many 5-loop master p-integrals
in  \cite{Georgoudis:2018olj}.

\section{\EQN{Big:no:pi} \BIG No-$\pi$ Theorem }

In this and next  Sections we will consider the  case of a generic one-charge theory, not  
necessarily QCD. We will not assume any specific properties of the corresponding 
$\beta$-function  and ADs, in particular $\beta_3^{\z_3}$ can have  non-zero value.

In fact,  the $\hG$-scheme suggests a strong generalization of  the no-$\pi$ theorem  
proven in Section \ref{interplay:generic:z34}. Indeed, the following  statement  is 
true. 

\vglue 0.2cm
\noindent
{\bf \BIG  No-$\pi$ Theorem } 
\\ Let $F$ be any $L$-loop massless correlator and all
$L$-loop p-integrals form  a $\pi$-safe class.  Then $F$ is
$\pi$-free in any (massless) renormalization scheme for which  corresponding   $\beta$-function 
and AD $\g$ are both $\pi$-free at least at the level of $L+1$  loops. 
\vglue 0.2cm

\newcommand{\hci}{\hat{c}_i}
\newcommand{\hbi}{\hat{b}_i}

\noindent
{\bf Proof }
\\
The theorem is obviously true for the $\hG$-scheme. Let  us  denote the  corresponding 
coupling constant as $\ha$, 
($\pi$-free!) RG functions as $\hbe(\hat{a})$ and $\hga(\ha)$, and  the very function $F$ as $\hF$.   
For  a different scheme  the  functions $F(a)$,  $\beta(a)$ and  $\g(a)$ 
are related via the  standard conversion formulas:
\bea
a &=&\ha\left(1 + \sum_{1 \le i \le L}{\hci}\,\ha^i\right),
\EQN{conv:a}
\\
F(a) &=& \left (1+ \sum_{1 \le i \le L} \hbi \,\ha^i\right) \hat{F}(\ha),
\EQN{conv:F}
\\
\be(a) &=&  \hbe(\hat{a}) + {\left(\sum_{1 \le i \le L}{i\,\hci}\,\ha^i\right)}\Bigg/{
\left (1+ \sum_{1 \le i \le L} \hci \,\ha^i\right)}
,
\EQN{conv:be}
\\
\g(a) &=&  \hga(\ha)  +
\hbe(\hat{a})\,\left(\sum_{1 \le i \le L}{i\,\hbi}\,\ha^i\right) 
\Bigg/
\left(1+ \sum_{1 \le i \le L} \hbi \,\ha^i\right)
\EQN{conv:ga}
{},
\eea
where its is understood that $\ha$ in eqs. (\ref{conv:F}--\ref{conv:ga})
is expressed in terms of  $a$ with  the use of the inverted   version of \re{conv:a}.

As relation \re{conv:a} connects two $\pi$-independent to order $(L+1)$ schemes  all coefficients
$c_i$ in it  should be $\pi$-free (see the comment after eq. \re{beta:pi:ind}). 

Thus, $\hbe(\ha(a))$ and $\hga(\ha(a))$ are $\pi$-free to order $(L+1)$. Finally, 
the requirement  that  
$
\gamma(a) 
\epi
 {\cal O}( a^{L+2}  )
$
and eq.~\re{conv:ga} mean
all coefficients  $b_i$ with $i=1,\dots L$ should also be $\pi$-free.

\section{ \EQN{6loops} Odd and even zetas in RG-functions up to 6 loops}

The $\hG$-scheme has some  remarkable  features. Indeed, one can see
just from its  definition that the corresponding ``hatted'' Green function, ADs
and $Z$-factors can be obtained from the normal (that is computed with the $G$-scheme)  by
very simple rules.

\begin{description}

\item [$\bullet$]
As a first  step we make  a formal replacement  of the coupling constant
 $a$ by  $\ha$ in every G-renormalized Green function, AD  and Z-factor we
want  to  transform to  the $\hG$-scheme.

\item[$\bullet$] 
Renormalized Green function $\hF(\ha)$  is obtained from $F(\ha)$ by setting to  zero {\em all}  even 
zetas  in the latter (both are assumed  as taken at $\ep=0$).  

\item[$\bullet$]
The same rule works for ADs and  $\beta$-functions.

\item[$\bullet$] 
If $Z$ is a ($G$-scheme)  renormalization constant then one should  not only 
nullify all even zetas in $Z(\ha)$ but also replace every odd zeta term in it  with its ``hatted'' 
counterpart.  
\end{description}

For future reference  we write below the connection   between  hatted and normal zeta for
the class of 5-loop p-integrals \cite{Georgoudis:2018olj}:
\begin{equation}\label{hatted:5L}
\begin{split}
                \hat{\zeta}_3 
                &\defas \zeta_3 + \frac{3 \epsilon}{2} \zeta_4 - \frac{5 \epsilon^3}{2} \zeta_6 + \frac{21 \epsilon^5}{2} \zeta_8\,, 
                \qquad
                \hat{\zeta}_5 
                \defas \zeta_5 + \frac{5 \epsilon}{2} \zeta_6 - \frac{35 \epsilon^3}{4} \zeta_8\,,
                \\
                \hat{\zeta}_7 
                &\defas \zeta_7 + \frac{7 \epsilon}{2} \zeta_8 \,,
                \qquad
                \hat{\varphi} 
                \defas \varphi -3\epsilon\, \zeta_4 \,\zeta_5 + \frac{5 \epsilon}{2} \zeta_3\, \zeta_6
                \qquad\text{and}\qquad
                \hat{\zeta}_9\defas \zeta_9,
\end{split}
\end{equation}
where        
\ice{$\varphi=\mzv{6,2}- \mzv{3,5} \approx -0.1868414$ }
\beq
\varphi \defas \frac{3}{5}\, \z_{5,3} + \z_3\, \z_5 -\frac{29}{20} \,\z_8 
= \mzv{6,2}- \mzv{3,5}
\approx -0.1868414
\EQN{varphi}
\eeq
and   
multiple zeta values are defined  as\footnote{Note, that the papers \cite{Kompaniets:2017yct,Georgoudis:2018olj}
use somewhat  different notation for  multiple zeta values: 
$\Bigr[\mzv{n_1,n_2} \Bigl]^{\mathrm{ our}} 
= \Bigr[\mzv{n_2,n_1} \Bigl]^{\text{\cite{Kompaniets:2017yct,Georgoudis:2018olj}}}
$.
The definition \re{mzv} is in agreement to the one employed, e.g. in 
\cite{Blumlein:2009cf,Broadhurst2013,Lee:2015eva,Schnetz:2016fhy}.  
}
\beq
\mzv{n_1,n_2} \defas \sum_{i > j > 0 } \frac{1}{i^{n_1} j^{n_2}}
\EQN{mzv}
{}.
\eeq

Once we know which objects appear at 5-loop p-integrals we can  construct a second Table
Бdisplaying the structure of 5-loop p-integrals and 6-loop ADs (in constructing the Table and in what  follows we 
assume that  {\em all} class of 5-loop p-integrals is $\pi$-safe \cite{Georgoudis:2018olj}).
Once again,  some  combinations including even zetas, namely,  $\z_4\z_5$ and $\z_3\z_6$
do not appear in   Table \ref{tab2} essentially  due to the $\pi$-safeness of 5-loop 
p-integrals as expressed by
\re{hatted:5L}. 

It is interesting to note that representation \re{hatted:5L} is in a sense
"too deep" for the  task of presentation of 5-loop p-integrals in the hatted
form. Indeed, a simple inspection of Table \ref{tab2} clearly shows that the
hatted form of a 5-loop p-integral can not depend on the last available terms
in $\ep$-expansions of $\hat{\z}_3$, $\hat{\z}_5$ and $\hat{\varphi}$ (we do not
consider the higher than $\ep^0$ terms in the $\ep$-expansion of 5-loop
p-integrals).  

\begin{center}
\begin{table}
\begin{center}
\begin{tabular}{ |c|c||c|c| } 
\hline
 L  & p-integrals  & L+1 &  Z 
\\
\hline
5  & 
$\z_3/\ep^3, \{\z_4, \z_5\}/\ep^2, \{\z_3^2, \z_6, \z_7\}/\ep$, 
&  6  %
&   $\z_3/\ep^4, \{\z_4, \z_5\}/\ep^3, \{\z_3^2, \z_6, \z_7 \xcancel{\z_3 \z_4}\}/\ep^2$, \\
\phantom{5}  & 
$\z_3\z_4$, $\z_8,\, \z_3\z_5,\, \varphi,\, \z_3^3,\, \z_9\ice{,\, \z_4\z_5,\, \z_3\z_6}$  
&  \phantom{6}  %
&  $\{  \z_3\z_4, \z_8,\, \z_3\z_5,\,  \varphi,\, \z_3^3,\, \z_9   \}/\ep$ \\
\hline
\end{tabular}
\caption{
\EQN{tab2}
\small 
The structure of 5-loop p-integrals (expanded in $\ep$ up to and including the
constant $\ep^0$ part)
and 6-loop  RCs.   The inverse power of $\ep$ stands for the {\em maximal} one in generic case;
in particular cases it might be less. 
}
\end{center}
\end{table}
\end{center}

Our next aim is to find explicitly the coefficients $c_i$ in the relation
between the charge RCs
$\hZa$ and $\Za$ in  $\hG$- and  $G$-schemes 
\beq
\ha =a\,\left(1 + \sum_{1 \le i \le L}{c_i}\,a^i\right)
\EQN{conv:ha}
{}.
\eeq
As the bare charge must not depend on  the  choice of the renormalization scheme   the 
coefficients $c_i$ are  fixed by requiring that 
\beq
\Za\,a = \hZ_a (\ha)\ha  
\EQN{master:ci}
{},
\eeq
where $\ha$ is expressed in terms of $a$ via  eq.~\re{conv:ha}.
A simple counting of powers of $a$ in \re{master:ci} shows that 
one can find  the coefficients $c_1 \dots c_6$ in terms of $\beta_1\dots\beta_6$.

For simplicity we start from the   case of 4 loops.
On  general grounds we can  write 
\beq 
\be = \be_1 a + \be_2 a^2 + 
(r_3  + \be_3^{\z_3}\,\z_3)\,a^3 +  (r_4  + \be^{\z_3}_4\,\z_3    + \be^{\z_4}_4\,\z_4         
 + \be^{\z_5}_4\,\z_5  )\,a^4       
{},
\eeq
where $r_i$ is   $\be_i$  with all zetas set to zero.
The corresponding  RCs  $Z_a$ and $\hZ_a$   read:
\begin{eqnarray}
Z_a &=& 1
+\frac{a\beta_1}{\ep}
+a^2\,\big(
\frac{1}{2\ep}\,\beta_2
+\frac{1}{\ep^2}\,\beta_1^2
\big)
+a^3\,\big(
\frac{1}{3\ep}\,(r_3
+\beta_3^{\z_3}\,\z_3)
+\frac{7}{6\ep^2}\,\beta_1\,\beta_2
+\frac{1}{\ep^3}\,\beta_1^3
\big)
\nonumber\\
&&+a^4\,\big(
\frac{1}{4\ep}\,(r_4
+\beta_4^{\z_3}\,\z_3
+\beta_4^{\z_4}\,\z_4
+\beta_4^{\z_5}\,\z_5)
+\frac{1}{\ep^2}\,(\frac{5}{6}\,\beta_1\,r_3
+\frac{5}{6}\,\beta_1\,\beta_3^{\z_3}\,\z_3
+\frac{3}{8}\,\beta_2^2)
\nonumber\\
&&
+\frac{23}{12\ep^3}\,\beta_1^2\,\beta_2
+\frac{1}{\ep^4}\,\beta_1^4
\big)
\end{eqnarray}
\ice{\begin{verbatim}
                                                          b3   b3z3 z3
                   2                        3             -- + -------
              2  b1     b2     a b1    3  b1    7 b1 b2   3       3
Out[15]= 1 + a  (--- + ----) + ---- + a  (--- + ------- + ------------) + 
g                   2   2 ep     ep          3        2         ep
                 ep                       ep     6 ep
 
                               2
                           3 b2    5 b1 b3   5 b1 b3z3 z3
           4        2      ----- + ------- + ------------
      4  b1    23 b1  b2     8        6           6
>    a  (--- + --------- + ------------------------------ + 
           4         3                    2
         ep     12 ep                   ep
 
        b4   b4z3 z3   b4z4 z4   b4z5 z5
        -- + ------- + ------- + -------
        4       4         4         4
>       --------------------------------)
                      ep
\end{verbatim}

\newpage
}
and
\begin{eqnarray}
\hat{Z}_a &=& 1
+\frac{\hat{a}}{\ep}\,\beta_1
+\hat{a}^2\,\big(
\frac{1}{2\ep}\,\beta_2
+\frac{1}{\ep^2}\,\beta_1^2
\big)
+\hat{a}^3\,\big(
\frac{1}{3\ep}\,(r_3
+\beta_3^{\z_3}\,\hat{\z}_3)
+\frac{7}{6\ep^2}\,\beta_1\,\beta_2
+\frac{1}{\ep^3}\,\beta_1^3
\big)
\nonumber\\
&&
+\hat{a}^4\,\big(
\frac{1}{4\ep}\,(r_4
+\beta_4^{\z_3}\,\hat{\z}_3
+\beta_4^{\z_5}\,\hat{\z}_5)
+\frac{1}{\ep^2}\,(
\frac{5}{6}\,\beta_1\,r_3
+\frac{5}{6}\,\beta_1\,\beta_3^{\z_3}\,\hat{\z}_3
+\frac{3}{8}\,\beta_2^2)
\nonumber\\
&&
+\frac{23}{12\ep^3}\,\beta_1^2\,\beta_2
+\frac{1}{\ep^4}\,\beta_1^4
\big)
{}.
\end{eqnarray}
\ice{
\begin{verbatim}


 
             4   4    3   3       4   2       2   2      3
             a  b1    a  b1    23 a  b1  b2   a  b1    7 a  b1 b2
Out[20]= 1 + ------ + ------ + ------------ + ------ + ---------- + 
                4        3             3         2           2
              ep       ep         12 ep        ep        6 ep
 
        4   2           2         4                       3
     3 a  b2    a b1   a  b2   5 a  b1 (b3 + b3z3 hz3)   a  (b3 + b3z3 hz3)
>    -------- + ---- + ----- + ----------------------- + ------------------ + 
          2      ep    2 ep                 2                   3 ep
      8 ep                              6 ep
 
      4
     a  (b4 + b4z3 hz3 + b4z5 hz5)
>    -----------------------------
                 4 ep
\end{verbatim}
}

Equation \re{master:ci} can be easily solved with the result 
\begin{eqnarray}
c_1&=&c_2=0,
\nonumber\\
c_3 &=& 
- \frac{1}{2}\,\beta_3^{\z_3}\,\z_4
+ \frac{5\ep^2}{6}\,\beta_3^{\z_3}\,\z_6 
- \frac{7\ep^4}{2}\,\beta_3^{\z_3}\,\z_8
\nonumber
{},
\\
c_4 &=&
 \frac{1}{4\ep}\,(
 \beta_4^{\z_4}
-\beta_1\,\beta_3^{\z_3} 
)\,\z_4 
-\frac{3}{8}\,\beta_4^{\z_3}\,\z_4 
- \frac{5}{8}\,\beta_4^{\z_5}\,\z_6 
\nonumber\\
&&
+ \frac{5\ep}{12}\,\beta_1\,\beta_3^{\z_3}\,\z_6 
+  \ep^2\,(
\frac{5}{8}\,\beta_4^{\z_3}\,\z_6 
+ \frac{35}{16}\,\beta_4^{\z_5}\,\z_8
)
- \frac{7\ep^3}{4}\,\beta_1\,\beta_3^{\z_3}\,\z_8 
- \frac{21\ep^4}{8}\,\beta_4^{\z_3}\,\z_8 
{}.
\end{eqnarray}
\ice{
\begin{verbatim}
c_1=c_2=0;

c_3 = -(b3z3*z4)/2 + (5*b3z3*ep^2*z6)/6 - (7*b3z3*ep^4*z8)/2

c_4 =

(-3*b4z3*z4)/8 + ((-(b1*b3z3)/4 + b4z4/4)*z4)/ep - (5*b4z5*z6)/8 + 
 (5*b1*b3z3*ep*z6)/12 - (7*b1*b3z3*ep^3*z8)/4 - (21*b4z3*ep^4*z8)/8 + 
 ep^2*((5*b4z3*z6)/8 + (35*b4z5*z8)/16)
\end{verbatim}
}
As the coefficients $c_i$  have to be finite  at $\ep \to 0$ we arrive at the  connection 
\beq
\beta_4^{\z_4} = \beta_1\, \beta_3^{\z_3}
\EQN{beta:4}
{},
\eeq
which  has been already  established in Section \ref{interplay:generic:z34} (see eq.~\re{identity:2:beta}). 
Repeating the same reasonings  
for 5 and 6 loops we arrive at exact relations which express {\em all} $\pi$-dependent terms in 
the coefficients  $\beta_5$ and $\beta_6$ in terms of $\beta_1$, $\beta_2$ and  contributions (proportional to 
odd zetas) to $\beta_3,\beta_4$ and $\beta_3,\beta_4,\beta_5$ correspondingly.  The relations read
\begin{eqnarray}
\beta_5^{\zeta_4} &=& \frac{1}{2} \beta_2\,\beta_3^{\zeta_3} + \frac{9}{8}\beta_1\,\beta_4^{\zeta_3},
\nonumber\\
\beta_5^{\zeta_6} &=& \frac{15}{8}\beta_1\,\beta_4^{\zeta_5},
\nonumber\\
\beta_5^{\zeta_3\zeta_4}&=&0,
\EQN{beta:5}
\eea
\bea
\beta_6^{\zeta_4} &=& \frac{3}{4}\beta_2\,\beta_4^{\zeta_3} + \frac{6}{5}\beta_1\,\beta_5^{\zeta_3}, 		    
\nonumber\\
\beta_6^{\zeta_6} &=& -\beta_1^3\,\beta_3^{\zeta_3} + \frac{5}{4}\beta_2\,\beta_4^{\zeta_5} + 2\,\beta_1\,\beta_5^{\zeta_5},
\nonumber\\
\beta_6^{\zeta_8} &=& \frac{14}{5}\beta_1\,\beta_5^{\zeta_7},
\nonumber\\
\beta_6^{\zeta_3\zeta_4} &=& \frac{12}{5}\beta_1\,\beta_5^{\zeta_3^2},		    
\nonumber\\
\beta_6^{\zeta_3\zeta_6} &=& 0,
\nonumber\\
\beta_6^{\zeta_4\zeta_5} &=& 0					    
\EQN{beta:6}
{}.                                        
\end{eqnarray}

In exactly the same way we derive similar constraints on a generic AD $\g$.
Indeed,  let $F_L(a,\lQ)$  is a p-function with evolution eq.~\re{G:RG:evol}
and $\hat{F}_L(\ha,\lQ)$ is its $\hG$-scheme counterpart.  
Both functions  should be finite at $\ep \to 0$ and meet   a conversion relation:
\beq
\hat{F} = \left( 1 + \sum_{1 \le i \le L} b_i \, a^i\right) F + {\cal O}(a^{L+1})
\EQN{F}
\eeq
or, 
equivalently, 
\beq
 \hZ(\ha) = \left( 1 + \sum_{1 \le i \le L} b_i \, a^i\right)\,Z(a) +   {\cal O}(a^{L+1})
{}.
\EQN{Z}
\eeq
The finiteness of the coefficients $b_i$ leads to the following  constraints
\ice{

}
\begin{eqnarray}
\gamma_4^{\zeta_4} &=&  \frac{3}{2}\gamma_3^{\zeta_3}\,\beta_1 - \frac{1}{2}\gamma_1\,\beta_3^{\zeta_3},
\EQN{gamma:4}
\\
\gamma_5^{\zeta_4} &=&  \frac{3}{2}\gamma_4^{\zeta_3}\,\beta_1 + \frac{3}{2}\gamma_3^{\zeta_3}\,\beta_2 - \gamma_2\,\beta_3^{\zeta_3} - \frac{3}{8}\gamma_1\,\beta_4^{\zeta_3},
\nonumber\\
\gamma_5^{\zeta_6} &=&  \frac{5}{2}\gamma_4^{\zeta_5}\,\beta_1 - \frac{5}{8}\gamma_1\,\beta_4^{\zeta_5},
\nonumber\\
\gamma_5^{\zeta_3\zeta_4} &=& 0,
\EQN{gamma:5}
\eea
\bea
\gamma_6^{\zeta_4} &=&  \frac{3}{2}\gamma_5^{\zeta_3}\,\beta_1 + \frac{3}{2}\gamma_4^{\zeta_3}\,\beta_2 
 + \frac{3}{2}\gamma_3^{\zeta_3}\,r_3 - 
 \frac{3}{2}(\gamma_3 - \gamma_3^{\z_3}\,\z_3)\,  \beta_3^{\zeta_3} - \frac{3}{4}\gamma_2\,\beta_4^{\zeta_3} 
- \frac{3}{10}\gamma_1\,\beta_5^{\zeta_3},
\nnb
\\
\gamma_6^{\zeta_6}  &=&  \frac{5}{2}\gamma_5^{\zeta_5}\,\beta_1 - \frac{5}{2}\gamma_3^{\zeta_3}\,\beta_1^3 + \frac{5}{2}\gamma_4^{\zeta_5}\,\beta_2 + 
          \frac{3}{2}\gamma_1\,\beta_1^2\,\beta_3^{\zeta_3} - \frac{5}{4}\gamma_2\,\beta_4^{\zeta_5} - \frac{1}{2}\gamma_1\,\beta_5^{\zeta_5},
\nonumber\\
\gamma_6^{\zeta_8}  &=&  \frac{7}{2}\gamma_5^{\zeta_7}\,\beta_1 - \frac{7}{10}\gamma_1\,\beta_5^{\zeta_7},
\nonumber\\
\gamma_6^{\zeta_3\zeta_4} &=&  3\,\gamma_5^{\zeta_3^2}\,\beta_1 - \frac{3}{5}\gamma_1\,\beta_5^{\zeta_3^2},
\nonumber\\
\gamma_6^{\zeta_3\zeta_6} &=&0,
\nonumber\\
\gamma_6^{\zeta_4\zeta_5} &=&0
{}.
\EQN{gamma:6}
\end{eqnarray}
Note, that if $\gamma=\beta$ then  relations (\ref{gamma:4}--\ref{gamma:6})  
become equivalent to 
(\ref{beta:4}--\ref{beta:6}) correspondingly.

We have successfully checked    all the constraints
above \ice{(\ref{beta:5} and \ref{gamma:6})}
for  a number of particular examples.

In QCD it was done for the 5-loop $\beta$-function and the quark mass
AD known from
\cite{Herzog:2017ohr,Luthe:2017ttg,Luthe:2016xec,Baikov:2017ujl} as well as
for all vertex and field ADs (taken in the  Landau gauge) computed in
\cite{Chetyrkin:2017bjc}.

At the level of 6 loops we have checked that the 
($\pi$-dependent contributions to) terms of order $n_f^5\alpha_s^6$ (for the $\beta$-functions) 
as well as   terms of order $n_f^5\alpha_s^6$ and  of order $n_f^4\alpha_s^6$ 
(the quark AD) computed in  \cite{Gracey:1996he,Ciuchini:1999cv,Ciuchini:1999wy} 
are in agreement with  relations (\ref{beta:6}) and 
(\ref{gamma:6}) respectively.

For the case of the  normal QCD with the $SU(3)$ gauge group our results
for $\pi$-dependent contributions to the 6-loop coefficients of the $\beta$-function
and the quark mass AD read (the known terms are  boxed)    
\begin{eqnarray}
\beta_6 &\epi& \boxed{\frac{608}{405}\,n_f^5\,\z_4}
+n_f^4\,(\frac{164792}{1215}\,\z_4
-\frac{1840}{27}\,\z_6)
+n_f^3\,(
-\frac{4173428}{405}\,\z_4
+\frac{1800280}{243}\,\z_6)
\nonumber\\&&
+n_f^2\,(\frac{68750632}{405}\,\z_4
-\frac{13834700}{81}\,\z_6)
+n_f\,(
-\frac{146487538}{135}\,\z_4
+\frac{40269130}{27}\,\z_6)
\nonumber\\&&
+99\,(44213\,\z_4
-64020\,\z_6)
{},
\end{eqnarray} 
\begin{eqnarray}
\gamma^m_6 &\epi& 
\boxed{
\frac{320}{243}\,n_f^5\,\z_4
+n_f^4\,(
-\frac{90368}{405}\,\z_4
+\frac{22400}{81}\,\z_6)
}
\nonumber\\&&
+n_f^3\,(
-\frac{92800}{27}\,\z_3\,\z_4
-\frac{2872156}{405}\,\z_4
+\frac{503360}{243}\,\z_6)
\nonumber\\&&
+n_f^2\,(\frac{661760}{9}\,\z_3\,\z_4
+\frac{155801234}{405}\,\z_4
-\frac{378577520}{729}\,\z_6
+\frac{12740000}{81}\,\z_8)
\nonumber\\&&
+n_f\,(
-\frac{1413280}{3}\,\z_3\,\z_4
-\frac{4187656168}{1215}\,\z_4
+\frac{5912758120}{729}\,\z_6
-\frac{96071360}{27}\,\z_8)
\nonumber\\&&
+3194400\,\z_3\,\z_4
+\frac{272688530}{81}\,\z_4
-\frac{6778602160}{243}\,\z_6
+15889720\,\z_8
.
\end{eqnarray} 
The  results for  generic gauge group are rather bulky, they can be found in
 Mathematica \cite{Mathematica} and FORM \cite{Vermaseren:2000nd}  ancillary 
files in the arxiv submission of this paper.

All 6-loop RG functions   known from 
\cite{Batkovich:2016jus,Schnetz:2016fhy,Kompaniets:2017yct} 
for the case of the $O(n)$ $\phi^4$ model  agree with  the constraints  
(provided that the transcendental constant   $\mzv{5,3}$ is replaced by
$\varphi$ according to relation \re{varphi}).

\section{\EQN{6loops:2} Odd and even zetas in p-functions up to 5 loops}

The knowledge  of the  coefficients $c_i$ and $b_i$ in the  relations
\re{conv:ha} and \re{F} allows to reconstruct   the full p-function $F$ from its  hatted version
$\hat{F}$. Indeed, let 
\beq
F(a) =  1 + \sum_{1\le i \le L} F_i\,a^i
{}.
\eeq
Then 
\beq
\hat{F}(\ha) =  1 + \sum_{1\le i \le L} \hat{F}_i\,\ha^i \ \ \mbox{with} \ \  \hat{F}_i = \left(F_i\right)|_{\pi=0}
{}.
\eeq

 At the limit of $\ep \to 0$ the coefficients $c_i$ and $b_i$  
defined  in the  previous Section read 
(the constraints (\ref{beta:4}--\ref{beta:6}) and (\ref{gamma:4}--\ref{gamma:6})
 are taken into account; the  results can be also  found ia computer-readable form  
in \cite{files}) 
\begin{eqnarray}
c_1&=&c_2=0,
\nonumber\\
c_3&=&
-\frac{1}{2}\,\beta_3^{\zeta_3}\,\zeta_4,
\nonumber\\
c_4&=&
-\frac{3}{8}\,\beta_4^{\zeta_3}\,\zeta_4
 - \frac{5}{8}\,\beta_4^{\zeta_5}\,\zeta_6,
\nonumber\\
c_5&=&
-\frac{3}{10}\,\beta_5^{\zeta_3}\,\zeta_4
 - \frac{3}{5}\,\beta_5^{\zeta_3^2}\,\zeta_3\,\zeta_4
 + \frac{1}{4}\,\beta_1^2\,\beta_3^{\zeta_3}\,\zeta_6 
 - \frac{1}{2}\,\beta_5^{\zeta_5}\,\zeta_6
 - \frac{7}{10}\,\beta_5^{\zeta_7}\,\zeta_8.
\EQN{ci}
\end{eqnarray}
and
\begin{eqnarray}
b_1&=&b_2=0,
\nonumber\\
b_3&=&-\frac{1}{2}\gamma_3^{\zeta_3}\,\zeta_4,
\nonumber\\
b_4&=&-\frac{3}{8}\,\gamma_4^{\zeta_3}\,\zeta_4 - \frac{5}{8}\gamma_4^{\zeta_5}\,\zeta_6,
\nonumber\\
b_5 &=&
       - \frac{3}{10}\,\gamma_5^{\zeta_3}\,\zeta_4
       - \frac{3}{5}\,\gamma_5^{\zeta_3^2}\,\zeta_4\,\zeta_3
       + (  
- \frac{1}{2}\,\gamma_5^{\zeta_5} 
+ \frac{1}{2}\,\beta_1^2\,\gamma_3^{\zeta_3} 
- \frac{1}{4}\,\beta_3^{\zeta_3}\,\beta_1\,\gamma_1 )\,\zeta_6
       - \frac{7}{10}\,\gamma_5^{\zeta_7}\,\zeta_8.
\EQN{bi}
\end{eqnarray}

\ice{
\begin{verbatim}
c_1=1

c_2=0

c_3=-(b3z3*z4)/2

c_4=(-3*b4z3*z4)/8 - (5*b4z5*z6)/8

c_5=(-3*b5z3*z4)/10 - (3*b5z32*z3*z4)/5 + (b1^2*b3z3*z6)/4 
     - (b5z5*z6)/2 -     (7*b5z7*z8)/10

c_5 =
       + z4 * (  - 3/10*b5z3 )

       + z4*z3 * (  - 3/5*b5z32 )

       + z6 * (  - 1/2*b5z5 + 1/4*b3z3*b1^2 )

       + z8 * (  - 7/10*b5z7 );

c_6 =
       + z4 * (  - 1/4*b6z3 )

       + z4*z3 * (  - 1/2*b6z32 )

       + z4^2 * (  - 3/10*b5z32*b1 + 1/2*b3z3^2 )

       + z5*z4 * ( 1/2*b6om )

       + z6 * (  - 5/12*b6z5 + 1/4*b4z3*b1^2 + 1/4*b3z3*b1*b2 )

       + z6*z3 * (  - 5/12*b6om )

       + z8 * ( 7/8*b4z5*b1^2 );



and

b_1=0

b_2=0

b_3=-(a3z3*z4)/2

b_4=(-3*a4z3*z4)/8 - (5*a4z5*z6)/8

b_5 =
       + z4 * (  - 3/10*a5z3 )

       + z4*z3 * (  - 3/5*a5z32 )

       + z6 * (  - 1/2*a5z5 + 1/2*b1^2*a3z3 - 1/4*b3z3*b1*a1 )

       + z8 * (  - 7/10*a5z7 );

b_6 =
       + z4 * (  - 1/4*a6z3 )

       + z4*z3 * (  - 1/2*a6z32 )

       + z4^2 * ( 1/8*a3z3^2 - 3/8*b1*a5z32 + 3/40*b5z32*a1 + 3/8*b3z3*a3z3 )

       + z5*z4 * ( 1/2*a6om )

       + z6 * (  - 5/12*a6z5 + 5/6*b1*b2*a3z3 + 5/12*b1^2*a4z3 - 1/6*b4z3*b1*
         a1 - 1/6*b3z3*b2*a1 - 5/12*b3z3*b1*a2 )

       + z6*z3 * (  - 5/12*a6om )

       + z8 * ( 35/24*b1^2*a4z5 - 7/12*b4z5*b1*a1 );


\end{verbatim}
}
Finally, we can now restore all $\pi$-dependent terms in $F_1-F_5$ from  eqs.~(\ref{F},\ref{ci})
and \re{bi} with the 
result:
\begin{eqnarray}
F_1 &=&\hat{F}_1, 
\EQN{F1}
\\
F_2 &=&\hat{F}_2,
\EQN{F2}
\\
F_3 &=&\hat{F}_3  +\frac{1}{2}\, \g_3^{\z_3} \, \z_4,
\EQN{F3}
\\
F_4 &=&\hat{F}_4
    + (  
 - \frac{1}{2}\,\hat{F}_1\,\beta_3^{\z_3} 
 + \frac{3}{8}\,\gamma_4^{\z_3} 
 + \frac{1}{2}\,\gamma_3^{\z_3}\,\hat{F}_1 )\,\z_4
    + \frac{5}{8}\,\gamma_4^{\z_5}\,\z_6,
\EQN{F4}
\\
F_5 &=&\hat{F}_5
 + (  
 - \beta_3^{\z_3}\,\hat{F}_2 
 - \frac{3}{8}\,\hat{F}_1\,\beta_4^{\z_3}
 + \frac{3}{10}\,\gamma_5^{\z_3}
 + \frac{3}{8}\,\gamma_4^{\z_3}\,\hat{F}_1
 + \frac{1}{2}\,\gamma_3^{\z_3}\,\hat{F}_2 )\,\z_4
       + \frac{3}{5}\,\gamma_5^{\z_3^2}\,\z_4\,\z_3
\nonumber\\
&&       + ( 
 \frac{1}{2}\,\gamma_5^{\z_5}
 + \frac{1}{4}\,\beta_3^{\z_3}\,\beta_1\,\gamma_1
 - \frac{5}{8}\,\hat{F}_1\,\beta_4^{\z_5}
 + \frac{5}{8}\,\hat{F}_1\,\gamma_4^{\z_5}
 - \frac{1}{2}\,\gamma_3^{\z_3}\,\beta_1^2 )\,\z_6
       + \frac{7}{10}\,\gamma_5^{\z_7}\,\z_8
{}.
\EQN{F5}
\end{eqnarray}

\ice{
Pavel! $a6om = \gamma_6^{\varphi}$ and  $hFi $ is $\hat{F}_i$
\bea
F_1 &=&\hat{F}_1, 
\\
F_2 &=&\hat{F}_2,
\\
F_3 &=&\hat{F}_3  \frac{1}{2}\, \g_3^{\z_3} \, \z_4
\eea 
\beq
F_4 =\hat{F}_4
\eeq
\begin{verbatim}
    + z4 * (  - 1/2*hF1*b3z3 + 3/8*a4z3 + 1/2*a3z3*hF1 )

    + z6 * ( 5/8*a4z5 );
\end{verbatim}
\beq
F_5 =\hat{F}_5
\eeq 
\begin{verbatim}
 + z4 * (  - b3z3*hF2 - 3/8*hF1*b4z3 + 3/10*a5z3 + 3/8*a4z3*hF1 + 1/2*
         a3z3*hF2 )

       + z4*z3 * ( 3/5*a5z32 )

       + z6 * ( 1/2*a5z5 + 1/4*b3z3*b1*a1 - 5/8*hF1*b4z5 + 5/8*hF1*a4z5 - 1/2*
         a3z3*b1^2 )

       + z8 * ( 7/10*a5z7 );
\end{verbatim}

\beq
F_6 =\hat{F}_6
\eeq 
\begin{verbatim}
       + z4 * (  - 3/4*b4z3*hF2 - 3/2*b3z3*hF3 - 3/10*hF1*b5z3 + 1/4*a6z3 + 3/
         10*a5z3*hF1 + 3/8*a4z3*hF2 + 1/2*a3z3*hF3 )

       + z4*z3 * ( 1/2*a6z32 - 3/5*hF1*b5z32 + 3/5*hF1*a5z32 )

       + z5*z4 * (  - 1/2*a6om )

       + z6 * ( 5/12*a6z5 - 5/4*hF2*b4z5 + 5/8*hF2*a4z5 + 1/6*b4z3*b1*a1 + 1/6
         *b3z3*a1*b2 + 5/12*b3z3*b1*a2 - 1/2*hF1*b5z5 + 1/2*hF1*a5z5 + 1/4*hF1
         *b3z3*b1*a1 + 1/4*hF1*b3z3*b1^2 - 5/12*a4z3*b1^2 - 5/6*a3z3*b1*b2 - 1/
         2*a3z3*hF1*b1^2 )

       + z6*z3 * ( 5/12*a6om )

       + z8 * ( 7/12*b1*a1*b4z5 - 35/24*b1^2*a4z5 - 7/80*b5z32*a1 + 7/16*a5z32
         *b1 - 7/10*hF1*b5z7 + 7/10*hF1*a5z7 - 7/16*a3z3*b3z3 + 7/48*a3z3^2 );

\end{verbatim}
}

We have checked that eqs. (\ref{F1}--\ref{F5}) correctly describe the $\pi$-dependent
contributions to all 11 4-loop p-functions from \cite{Ruijl:2017eht} as well
as their 5-loop scale invariant versions constructed according to the
definition \re{dG1}  with the use of 5-loop ADs from
\cite{Chetyrkin:2017bjc}.

\section{Discussion and Conclusions}

We have  clarified and proved the no-$\pi$ theorem for all one-scale RG-invariant Euclidean
correlators $\hat{F}^{si}_5$ first suggested in \cite{Jamin:2017mul}. 
The  theorem is extended to  a case of generic   Euclidean correlators with
arbitrary high loop number. 

We have found many new identities relating contributions proportional to odd and even zetas
in generic \MSbar\  ADs and $\beta$-functions.
The  new identities allow to reconstruct all  $\pi$-dependent terms in a RG function at the  level 
$L$  loops  in terms of  the same  function and the  $\beta$-function
taken at loop level $(L-1)$ and less with $L =4,5$ and $6$.
We have explicitly  elaborated the corresponing predictions for $L=4,5$ and $6$ 
and  found  full agreement of our results  with calculated results where the latter are available.

For the case of the $O(n)$ $\phi^4$ theory there exist since recently results
for the corresponding ADs at the 7 loop level
\cite{Schnetz:2016fhy}. In fact, we have extended our formulas to this case
and successfully reproduced  all the   $\pi$-dependent terms
in 7-loop ADs  and the 6-loop propagator of the scalar  field.
We are planning   to publish a detailed account of these results  in  the near  future.

Finally, the use of the $\Ghat$-scheme is not limited to one-charge models
or/and to the Landau gauge fixing.  It is clear that the derivation of
Sections \ref{6loops} and \ref{6loops:2} can be straightforwardly extended on
a general case of a QFT model with a few coupling constants considered in a
generic covariant gauge. More loops are also not a problem {\em provided}
corresponding generalizations of representation \re{hatted:5L} can be
constructed.

Both authors are indebted to J.H. K\"uhn for careful reading the manuscript
and a lot of good advice.

We are very grateful to J. Gracey for his help in extracting 6-loop terms from generic
results of \cite{Gracey:1996he,Ciuchini:1999cv,Ciuchini:1999wy} and to
O.~Schnetz for providing us with his results for the 7-loop ADs and the bare
6-loop propagator in the $O(n)$ $\phi^4$ model.

The second author thanks  J.~Davies, M.~Jamin, B.~ Kniehl,
M.~Kompaniets, E.~Panzer, A.~Pikelner and A.~Vogt for useful discussions.

\ice{
\section{Discussion}

The use of the $\Ghat$-scheme is not limited  to one-charge models or/and  to the Landau gauge fixing.
It is  clear that   the derivation of   Section \ref{6loops} 
can be straightforwardly generalized for a general case of a  QFT model with a few coupling constants 
considered in a generic covariant gauge. More loops are also not  a problem  {\em provided}
corresponding generalizations of relations \re{hatted:5L} can be constructed. 
}


The work of P.A.~Baikov is supported in part by the 
grant RFBR 17-02-00175A of the Russian Foundation for Basic Research.
The work by K. G. Chetykin was supported by 
the German Federal Ministry for Education and Research BMBF
through Grant  No. \ice{05H2015  and}05H15GUCC1.
\vspace{4mm}

\noindent
{\bf Note Added }
\\
After completing our work we have been  informed by E.~Panzer that he 
independently has found  a way to reconstruct $\pi$-dependent   terms in RG
functions of the  $O(n)$ $\phi^4$ model.


\providecommand{\href}[2]{#2}\begingroup\raggedright\endgroup

\ed

\providecommand{\href}[2]{#2}\begingroup\raggedright\endgroup

\ed

\ed

The second way to get scale-invariant 

\beq
\hat{G}^{\mathrm si}_n = 
\EQN{SI:def}
\eeq

\beq
\frac{m(\mu)}{m(\mu_0)} = \frac{c(a_s(\mu))}{c(a_s(\mu_0))}, \  \ \ 
c(x) = \mathrm{exp}\Biggl\{ \int {d x'} \frac{\g_m(x'}{\beta(x')} \Biggr\} 
{},
\label{cfun:1}
\eeq

\bga
c(x) =  (x)^{\bar{\g_0}} \Bigg\{ 1 + d_1 x + (d_1^2/2 + d_2) \,x^2 
\\          +  (d_1^3/6 + d_1 d_2 + d_3)\,  {x^3} \nonumber 
 +         (d_1^4/24 + d_1^2 d_2/2 
\\
+ d_2^2/2+ d_1 d_3 + d_4)
\,{x^4} + {\cal O}(x^5) \Bigg\}
\label{cfun:2}
{},
\end{gather}

\begin{flalign}
& d_1 = -\bar{\beta}_1 \, \bar{\gamma}_0 + \bar{\gamma}_1 {},&
\\
&d_2 = \bar{\beta}_1^2 \, \bar{\gamma}_0/2 - \bar{\beta}_2 \, \bar{\gamma}_0/2 - \bar{\beta}_1 \, \bar{\gamma}_1/2 + \bar{\gamma}
_2/2
{}, &
\\
&d_3 = -\bar{\beta}_1^3 \, \bar{\gamma}_0/3 + 2 \, \bar{\beta}_1 \, \bar{\beta}_2 \, \bar{\gamma}_0/3 - \bar{\beta}_3 \, 
\bar{\gamma}_0/3
\\
& \hspace{11mm}
 + \bar{\beta}_1^2 \, \bar{\gamma}_1/3 
          -\bar{\beta}_2 \, \bar{\gamma}_1/3 - \bar{\beta}_1 \, \bar{\gamma}_2/3 + \bar{\gamma}_3/3
{},
\nnb
\\
&d_4 = \bar{\beta}_1^4 \, \bar{\gamma}_0/4 - 3 \, \bar{\beta}_1^2 \, \bar{\beta}_2 \, \bar{\gamma}_0/4 + \bar{\beta}_2^2 \, \bar{
\gamma}_0/4 
\label{d4}
\\
& \hspace{18mm} + \bar{\beta}_1 \, \bar{\beta}_3 \, \bar{\gamma}_0/2 
    -  \bar{\beta}_4 \, \bar{\gamma}_0/4 - \bar{\beta}_1^3 \, \bar{\gamma}_1/4 
\nonumber
   \\ 
   & \hspace{18mm} + \bar{\beta}_1 \,  \bar{\beta}_2 \, \bar{\gamma}_1/2 - \bar{\beta}_3 \, \bar{\gamma}_1/4 +   \bar{\beta}_1^2 \
, \bar{\gamma}_2/4 
\nnb
\\
& \hspace{18mm}
- \bar{\beta}_2 \, \bar{\gamma}_2/4 
    - \bar{\beta}_1 \, \bar{\gamma}_3/4 + \bar{\gamma}_4/4 {}.
\nnb
&
\end{flalign}

Here 
$\bar{\g_i} = (\g_m)_i/\beta_0$, $\bar{\beta}_i = \beta_i/\beta_0$   
 and 
\[\beta(a_s)  =-\sum_{i \ge 0} \,\beta_i \, a_s^{i+2} = -\beta_0 \left\{\sum_{i \ge 0} \,\bar{\beta_i}\,  a_s^{i+2}
\right\}
\]